\documentclass[12pt]{iopart}

\usepackage{mathabx}
\usepackage{graphicx}

\begin{document}

\title{Flat Electron Bands with Bad Valley Quantum Numbers in Twisted Bi-Layer Graphene}

\author{J.P. Rodriguez}

\address{Department of Physics and Astronomy, California State University,
 Los Angeles, California 90032, USA}
\ead{jrodrig@calstatela.edu}
\vspace{10pt}
\begin{indented}
\item[]August 2017
\end{indented}

\begin{abstract}
We compute the energy spectrum of a nearest-neighbor electron hopping model
for bi-layer graphene at commensurate twist angles.
Specifically, we focus on the simplest bi-layer lattices, with moir\'e patterns
that have no subcells.
The electron hopping hamiltonian is analyzed in momentum space,
 both by degenerate perturbation theory and by exact numerical calculation.
We find that the degeneracy in energy along the edge of the moir\'e Brillouin zone
due to the two valley quantum numbers is noticeably broken in the flat central bands
at the magic twist angle.  A mechanism for the appearance of flat central bands themselves
 at the magic twist angle is also revealed.
It is due to maximal level repulsion. 
 The mechanism  relies on the assumption that the phase factor for the AA
hamiltonian matrix element for inter-graphene-sheet hopping at the middle of the edge
of the moir\'e Brillouin zone has a phase equal to half the twist angle.
This assumption is confirmed in the case of uniform AA hopping in between the two sheets of graphene,
in the limit of large moir\'e unit cells.
\end{abstract}

%
%
%
\maketitle
%
%

\section{Introduction}
It is well known that electron-electron correlations can lead to insulating behavior in systems with unfilled bands.
This is the case, for example, in copper-oxides,
where high-temperature superconductivity emerges
after doping the correlated (Mott) insulator state with holes.
Similar behavior has been observed recently in bi-layer graphene at a magic twist angle of $\theta = 1.08^{\circ}$.
Superconductivity has been observed at a hole doping
 just beyond $\nu = -2$ electrons per moir\'e unit cell\cite{cao_tblg_nature_18a}.
And a correlated insulator state has been observed at $\nu = -2$ electrons per moir\'e unit cell,
as well as at $\nu = 2$ \cite{cao_tblg_nature_18b}.
This observation of superconductivity upon hole-doping the half-filled hole bands in
bi-layer graphene at the magic twist angle
was quickly reproduced by other groups\cite{lu_efetov_nature_19}\cite{saito_tblg_ntr_phys_20}.

That flat electron bands in twisted bi-layer graphene exist at the magic angle
was predicted theoretically a decade before
the experimental discovery noted above\cite{lopes_prl_07}\cite{suarez_prb_10}\cite{bistritzer_macdonald_11}.
In particular, these numerical calculations found that the velocity of the Dirac cones 
near charge neutrality is strongly renormalized down at the magic twist angle.
A fundamental understanding of this effect remains unkown, however.

In this paper, we study the electronic properties of bi-layer graphene at commensurate twist angles
on the basis of a nearest-neighbor electron hopping model.
We primarily focus on commensurate twists that produce periodic moir\'e patterns
with no subcells.
These result in the smallest number of carbon atoms per true moir\'e unit cell.
Degenerate perturbation theory reveals inter-valley mixing at the corners of the moir\'e Brillouin zone.
Direct calculation finds, however, that the corresponding matrix element is small by a factor
of one-over a quarter of the number of carbon atoms per true moir\'e unit cell.
Importantly, this result agrees with the popular twist construction of the reciprocal lattice
introduced by Bistritzer and MacDonald\cite{bistritzer_macdonald_11}, which neglects such transitions.
Both degenerate perturbation theory and exact numerical calculation find that valley denegenracy 
of the energy spectrum is broken near the corners of the moir\'e Brillouin zone.
Specifically,
exact numerical calculation reveals that the breaking of valley degeneracy
 becomes apparent in the dispersion of  the flat central energy bands at the magic twist angle
along the edge of
the moir\'e Brillouin zone. 

Following the Bistritzer-MacDonald twist construction\cite{bistritzer_macdonald_11}, on the other hand,
we apply degenerate perturbation theory at the middle of the edge of the moir\'e Brillouin zone.
We find that flat bands exist at the magic twist angle because of maximal level repulsion there.
This result relies on the assumption that the phase factor of
the corresponding inter-valley matrix element between A sites there 
has a phase equal to half the twist angle.
This assumption is corroborated in the case of uniform nearest-neighbor electron hopping
in between the two sheets of graphene, in the limit of large moir\'e unit cells.

The paper is organized as follows.  Section 2 describes the nature of
the lattice of bi-layer graphene at commensurate twist angles,
while the nature of the corresponding reciprocal lattice, which is dual to former,
is described in section 3.  The nearest-neighbor hopping hamiltonian 
is analyzed in section 4, where the flat-band effect is revealed.
Last, exact dispersions of the flat bands at the magic twist angle are computed numerically
in section 5.

\section{Bi-Layer Honeycomb Lattices at Commensurate Twist Angles\label{cmmnsrt_tblg}}
\begin{figure}
\includegraphics[scale=1.00, angle=0]{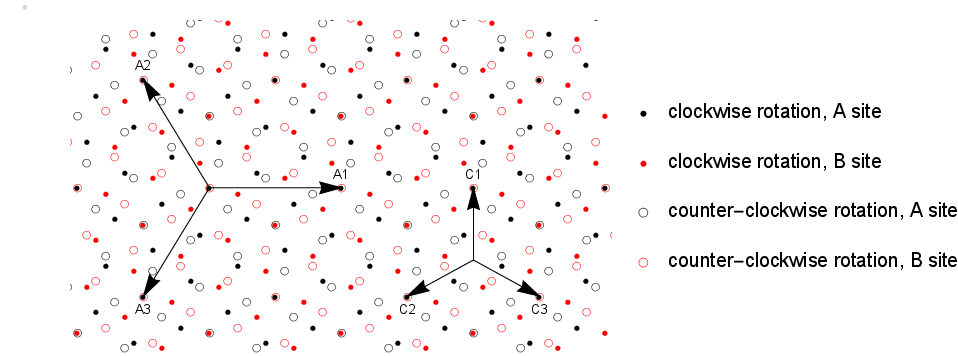}
\caption{Bi-layer graphene at the largest commensurate twist angle: $\theta = 21.8^{\circ}$.
It has integer parameters $m=2$ and $n=1$, with $l=7$ A(B) sites per moir\'{e} unit cell,
per sheet of graphene. [See Eqs. (\ref{sine}) and (\ref{l_M}).]}
\label{cmmnsrt_tblg_7}
\end{figure}

Consider bi-layer graphene with no twist.  
It exhibits Bernal stacking, with A sites on top of B sites, and vice versa.
Figure \ref{cmmnsrt_tblg_7} shows a commensurate bi-layer that twists about such Bernal stacks,
with the maximum twist angle of $\theta = 21.8^\circ$.
It notably shows three types of twist centers: Bernal stacks with A sites on top of B sites,
and vice versa,  and twist centers with rings of nearest-neighbor A sites in between graphene sheets
and nearest-neighbor B sites in between graphene sheets.  We will show below that an infinite series of such
commensurate twisted bi-layer graphene lattices exist with smaller twist angles that approach zero.

We start from bi-layer graphene in the Bernal-stacked form, with {\it no} twist.  
The (redundant) primitive vectors among the A(B) sites are:
${\bi a}_1 = a_{\triangle} {\hat {\bi y}}$, 
${\bi a}_2 = a_{\triangle} (-{\sqrt{3}\over 2}{\hat {\bi x}}-{1\over 2}{\hat {\bi y}})$, and
${\bi a}_3 = a_{\triangle} ({\sqrt{3}\over 2}{\hat {\bi x}}-{1\over 2}{\hat {\bi y}})$.
Here, $a_{\triangle} = \sqrt{3} a$ is the lattice constant, 
where $a$ is the carbon-carbon separation in graphene.
Now let $m$ and $n$ be positive integers that are relatively prime, with $m > n$,
and consider the following pair of displacements
 from a Bernal stacking\cite{suarez_prb_10}\cite{campanera_prb_07}\cite{trambly_nano_lett_10}:
${\bi A}_1(+) = -m {\bi a}_2 + n {\bi a}_3$ from the A site of the stack,
${\bi A}_1(-) = -n {\bi a}_2 + m {\bi a}_3$ from the B site of the stack.
They are related to each other through a reflection about the $x$ axis.
We can then create a commensurate lattice
by twisting the graphene sheet that contains the A site of the stack
 clockwise by an appropriate angle $\alpha$, and
by twisting the other graphene sheet that contains the B site of the stack
 counterclockwise by the same angle $\alpha$,
such that the two displacements coincide at a new Bernal stack.
Because the difference in the two displacement vectors is
${\bi A}_1(+)-{\bi A}_1(-) = (m-n) {\bi a}_1$,
while the magnitude of each displacement vector is
$|{\bi A}_1(\pm)| = (m^2 + m n + n^2)^{1/2} a_{\triangle}$,
we have that the equal and opposite twist angle $\alpha$ satisfies
\begin{equation}
2 \sin \alpha = (m-n)/\sqrt{l} ,
\label{sine}
\end{equation}
where
\begin{equation}
l = m^2 + m n + n^2 
\label{l_M}
\end{equation}
is the number of A(B) sites per {\it true} moir\'{e} unit cell, per sheet of graphene.
Upon such an equal and opposite twist, a triangular lattice of Bernal stacks is achieved,
with (redundant) moir\'{e} primitive vectors
%
\begin{eqnarray}
{\bi A}_1 = \sqrt{l}\, a_{\triangle} {\hat{\bi x}}, \quad
{\bi A}_2 = \sqrt{l}\, a_{\triangle} \Biggl(-{1\over 2} {\hat{\bi x}} + {\sqrt{3}\over 2} {\hat{\bi y}}\Biggr), \quad
{\bi A}_3 = \sqrt{l}\, a_{\triangle} \Biggl(-{1\over 2} {\hat{\bi x}} - {\sqrt{3}\over 2} {\hat{\bi y}}\Biggr), \nonumber \\
\label{A_i}
\end{eqnarray}
%
%
Repeating the previous, but swapping the A and B sites of the Bernal stacks,
results in another moir\'{e} triangular lattice.
We will show below that the two sets of Bernal stacks form a honeycomb lattice
if $m-n$ is not a multiple of $3$.

Again consider untwisted bi-layer graphene, but with the less common AA stacking instead.
Designate a center of one  honeycomb cell as a new twist center.
From that center, the A sites of the honeycomb lattice have (redundant) basis vectors:
${\bi c}_1 = a {\hat {\bi x}}$,
${\bi c}_2 = a (-{1\over 2}{\hat {\bi x}}+{\sqrt{3}\over 2}{\hat {\bi y}})$, and
${\bi c}_3 = a (-{1\over 2}{\hat {\bi x}}-{\sqrt{3}\over 2}{\hat {\bi y}})$.
Consider next an A(B) site in one sheet of graphene that is displaced from the designated twist center by
${\bi C}_1(+) = m {\bi c}_2 - n {\bi c}_3$, 
and a B(A) site in the other sheet of graphene that is displaced from the designated twist center by
${\bi C}_1(-) = n {\bi c}_2 - m {\bi c}_3$.
The magnitude of each displacement vector is
$|{\bi C}_1(\pm)| = (m^2 + m n + n^2)^{1/2} a$.
Recall now that
given a general displacement from the center
$p {\bi c}_2 + q {\bi c}_3$,
where $p$ and $q$ are integers,
then it is an A site for $(p+q)$ mod $3 = 1$,
a B site for $(p+q)$ mod $3 = 2$, and a center of a honeycomb cell
for $(p+q)$ mod $3 = 0$.
Yet the difference of the two displacements above is
${\bi C}_1(-)-{\bi C}_1(+) = (m-n) {\bi c}_1$.
We then conclude that twisting the graphene sheet with 
the A(B) displacement ${\bi C}_1(+)$ clockwise by the angle (\ref{sine}) $\alpha$,
and twisting the graphene sheet 
with the B(A) displacement ${\bi C}_1(-)$ counterclockwise by the angle $\alpha$
results in a Bernal stack if $m-n$ is {\it not} a multiple of $3$.
Such a twist applied to the displacements $-{\bi C}_1(\pm)$ results in another Bernal stack,
but with the A site and B site swapped.  The twist of each graphene sheet clockwise and counterclockwise
by the angle $\alpha$ thereby results in a moir\'{e} honeycomb of Bernal stacks
when $m-n$ is not a multiple of 3.
The displacement vectors of the Bernal stacks
of the resulting twisted bi-layer graphene lattice
from the twisted AA stacking center above are
%
\begin{eqnarray}
{\bi C}_1 = \sqrt{l}\, a  {\hat {\bi y}}, \quad
{\bi C}_2 = \sqrt{l}\, a \Biggl(-{\sqrt{3}\over 2}{\hat {\bi x}}-{1\over 2}{\hat {\bi y}}\Biggr), \quad
{\bi C}_3 = \sqrt{l}\, a \Biggl({\sqrt{3}\over 2}{\hat {\bi x}}-{1\over 2}{\hat {\bi y}}\Biggr). \nonumber \\
\label{C_i}
\end{eqnarray}
%
They are shown in Fig. \ref{cmmnsrt_tblg_7}.
Last, the case when $(m-n)$ mod $3 = 0$ coincides with a triangular lattice of Bernal stacks of the same type:
for example,
all A site of graphene sheet rotated counterclockwise over B site of graphene sheet rotated clockwise,
or vice versa.

\begin{figure}
\includegraphics[scale=1.00, angle=0]{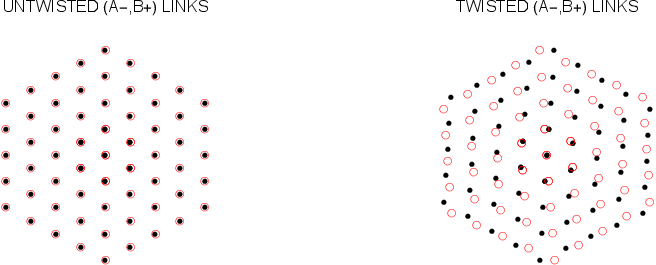}
\caption{Nearest neighbor links within a moir\'{e} unit cell
 between A sites in clockwise-rotated sheet of graphene (black dot)
and B sites in counterclockwise-rotated sheet of graphene (red circle)
for the case of $n=4$ consecutive hexagonal shells,
plus the twist center ($m=5$). 
It corresponds to $l=61$ A(B) sites in each moir\'e unit cell per sheet of graphene.
The commensurate twist angle is equal to $\theta = 7.3^{\circ}$.}
\label{twist_A_B}
\end{figure}

We shall next determine the nearest-neighbors between two sheets of graphene 
upon a commensurate twist, as just described.
Let us confine ourselves hearafter to the case where $m = n+1$,
which we name the Periodic Moir\'e Pattern line.
The true Wigner-Seitz unit cell has no moir\'e subcells in such case.
(Cf. Fig. \ref{moire_pattern_109}.)
The number of A(B) sites per graphene sheet, per moir\'e unit cell,
(\ref{l_M}) is then equal to 
$l = 1 + 3 n (n+1)$ in such case.  If we start from a Bernal stack in untwisted bi-layer graphene,
the moir\'{e} unit cell then corresponds to $n$ consecutive hexagonal shells of A and B sites in each
respective sheet of graphene.  Adding the twist center yields the 
total number of nearest-neigbor links inside the moir\'e unit cell
between A sites in the minus-rotated sheet of graphene and B sites in the plus-rotated sheet, $l$.
This is shown in Fig. \ref{twist_A_B} for the case $n=4$ and $m=5$,
which has  $l = 61$ A(B) sites in each moir\'{e} unit cell per sheet of graphene.
The sites corresponding to twisted bi-layer graphene are achieved by twisting the respective sheets
of graphene clockwise and counterclockwise by the angle (\ref{sine}) $\alpha$
about the Bernal stack at the center of the hexagonal shells.
See, again, Fig. \ref{twist_A_B}.

\begin{figure}
\includegraphics[scale=1.00, angle=0]{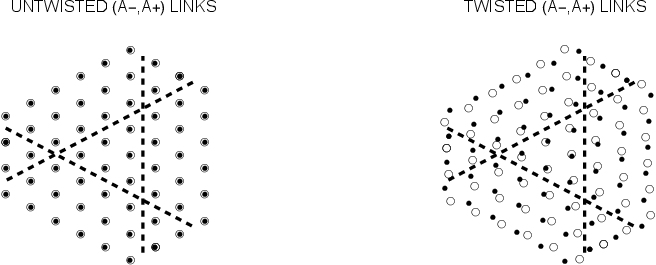}
\caption{Nearest neighbor links within a moir\'{e} unit cell
 between A sites in clockwise-rotated sheet of graphene (black dot)
and A sites in counterclockwise-rotated sheet of graphene (black circle).
The commensurate twist angle is equal to $\theta = 7.3^{\circ}$.
It corresponds to integer parameters $m=5$ and $n=4$,
with $l = 61$ A(B) sites per moir\'e unit cell, per sheet of graphene.}
\label{twist_A_A}
\end{figure}

Nearest-neighbor A sites between graphene sheets are shown in Fig. \ref{twist_A_A}.
The twist center is located at the center of the super-honeycomb cells formed by the Bernal stacks.
(See Fig. \ref{cmmnsrt_tblg_7}.)
As in the previous case, there exist $n$ hexagonal shells, with the exception
that the smallest shell forms a triangle.  There is no A site at the twist center, however.
And because inter-graphene-sheet links with an A site at a Bernal stack are excluded,
the number of such nearest-neighbor links inside a moir\'e unit cell is
$l-1 = 3 n (n+1)$.  Direct summation of the number of such A-A matrix elements
inside a moir\'e unit cell confirms this result.  (See \ref{ppndx_c_hmltn}.)
Last, the corresponding B sites are obtained by simple space inversion of the A sites about the twist center.

\section{Bloch Theorem}
Below, we shall construct the hamiltonian for an electron hopping in between nearest neighbors
in twisted bi-layer graphene at commensurate twist angles, e.g. Fig. \ref{cmmnsrt_tblg_7},
but in momentum space.

\begin{figure}
\includegraphics[scale=1.00, angle=0]{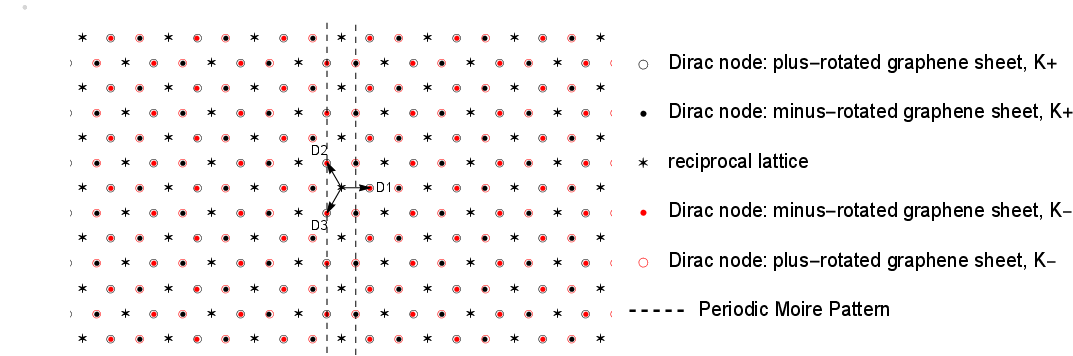}
\caption{Moir\'{e} reciprocal lattice ($*$) of bi-layer graphene
and the Dirac nodes corresponding to
the minus-rotated ($\cdot$) and plus-rotated ($\circ$) sheets of graphene
at commensurate twist angles.}
\label{rcprcl_lttc_a}
\end{figure}

\subsection{Momentum Space}
The (redundant) primitive vectors of the moir\'{e} reciprocal lattice can be obtained from the primitive vectors of the moir\'e lattice (\ref{A_i}),
and they are
\begin{equation}
{\bi B}_1 = {2\pi\over{\sqrt{l}\, d}} {\hat{\bi y}}, \quad
{\bi B}_2 = {2\pi\over{\sqrt{l}\, d}} \Biggl(-{\sqrt{3}\over 2} {\hat{\bi x}} - {1\over 2} {\hat{\bi y}}\Biggr), \quad
{\bi B}_3 = {2\pi\over{\sqrt{l}\, d}} \Biggl({\sqrt{3}\over 2} {\hat{\bi x}} - {1\over 2} {\hat{\bi y}}\Biggr),
\label{B_i}
\end{equation}
where 
$d = {\sqrt{3}\over 2} a_{\triangle} = {3\over 2} a$ 
is the shortest distance in between Bragg planes of A(B) sites in graphene.
Figure \ref{rcprcl_lttc_a} shows the resulting reciprocal lattice.
It is well known that momentum space is {\it inverted} by comparison with real space.
For example, by comparison with the moir\'e primitive vectors (\ref{A_i})
 ${\bi A}_1$, ${\bi A}_2$, and ${\bi A}_3$,
\begin{equation}
{\bi b}_1(+) = -m {\bi B}_2 + n {\bi B}_3 \quad {\rm and} \quad {\bi b}_1 (-) = -n {\bi B}_2 + m {\bi B}_3
\label{b_1}
\end{equation}
give the first primitive vectors of the plus-rotated and minus-rotated 
reciprocal lattices of the respective sheets of graphene.
Their un-rotated counterparts are
\begin{equation}
{\bi b}_1 = {2\pi\over{d}} {\hat{\bi x}}, \quad
{\bi b}_2 = {2\pi\over{d}} \Biggl(-{1\over 2} {\hat{\bi x}} + {\sqrt{3}\over 2} {\hat{\bi y}}\Biggr), \quad
{\bi b}_3 = {2\pi\over{d}} \Biggl(-{1\over 2} {\hat{\bi x}} - {\sqrt{3}\over 2} {\hat{\bi y}}\Biggr).
\label{b_i}
\end{equation}
%
Likewise,
by comparison with the displacement vectors ${\bi C}_1$,  ${\bi C}_2$, and  ${\bi C}_3$, in twisted bi-layer graphene
(Fig. \ref{cmmnsrt_tblg_7}),
the Dirac nodes in
the counter-clockwise-rotated sheet of graphene
and in the clockwise-rotated sheet of graphene
 are related to moir\'{e} Dirac nodes by
\begin{equation}
{\bi d}_1 (+) = m {\bi D}_2 - n {\bi D}_3 \quad {\rm and} \quad {\bi d}_1 (-) = n {\bi D}_2 - m {\bi D}_3 .
\label{d1}
\end{equation}
Here,
%
\begin{equation}
{\bi D}_1 = {K\over{\sqrt{l}}} {\hat{\bi x}}, \quad
{\bi D}_2 = {K\over{\sqrt{l}}} \Biggl(-{1\over 2} {\hat{\bi x}} + {\sqrt{3}\over 2} {\hat{\bi y}}\Biggr), \quad
{\bi D}_3 = {K\over{\sqrt{l}}} \Biggl(-{1\over 2} {\hat{\bi x}} - {\sqrt{3}\over 2} {\hat{\bi y}}\Biggr),
\label{D_i}
\end{equation}
are moir\'{e} Dirac nodes that are closest to the zero reciprocal lattice vector,
as shown in Fig. \ref{rcprcl_lttc_a},
while
\begin{equation}
{\bi d}_1 = K {\hat{\bi y}}, \quad
{\bi d}_2 = K \Biggl(-{\sqrt{3}\over 2} {\hat{\bi x}} - {1\over 2} {\hat{\bi y}}\Biggr), \quad
{\bi d}_3 = K \Biggl({\sqrt{3}\over 2} {\hat{\bi x}} - {1\over 2} {\hat{\bi y}}\Biggr),
\label{d_i}
\end{equation}
are the  Dirac nodes of an  un-rotated sheet of graphene.
Above, $K = 4\pi / 3 a_{\triangle}$
is the lattice constant of the honeycomb of such Dirac nodes.

Now notice by (\ref{d1}) that the twist in the location of the Dirac node at ${\bi d}_1$ results in the difference
${\bi d}_1 (-) - {\bi d}_1 (+) = (m-n) {\bi D}_1$.
By  expression (\ref{D_i}) for ${\bi D}_1$,
the twist angle $2 \sin\alpha = |{\bi d}_1 (-) - {\bi d}_1 (+)|/K$
thereby coincides with the formula (\ref{sine}) obtained in real space.
At integer parameters $m-n = 1$,
 the reciprocal lattice of the bi-layer graphene at commensurate twist
coincides with the twist construction of the reciprocal lattice
introduced by Bistritzer and MacDonald \cite{bistritzer_macdonald_11}.
In that case, for example,
the lattice constant of the moir\'e Dirac nodes shown in Fig. \ref{rcprcl_lttc_a}
is  related to the lattice constant of the Dirac nodes in an isolated sheet of graphene by
\begin{equation}
K_M = 2 (\sin \alpha) K.
\label{K_M}
\end{equation}
The resulting moir\'e pattern is thereby periodic\cite{lopes_prl_07}.
Henceforth, we shall therefore call the constraint $m-n = 1$ the Periodic Moir\'e Pattern line.
For integer parameters along the Periodic Moir\'e Pattern line, 
 the Dirac nodes ${\bi d}_1 (+)$ and ${\bi d}_1 (-)$
for the plus-rotated and minus-rotated sheets of graphene,  respectively, lie along the 
left dashed line and the right dashed line that are shown in Fig. \ref{rcprcl_lttc_a}.
Figure \ref{rcprcl_lttc_b} shows the moir\'{e} reciprocal lattice plus the Dirac nodes corresponding
to each sheet of graphene in the case of integer parameters $m=5$  and $n=4$.
It has $l = 61$ A(B) sites per sheet of graphene in a moir\'{e} unit cell.
Notice that the first Brillouin zone for each sheet of graphene
contain  $n=4$ consecutive hexagonal shells
of moir\'{e} reciprocal lattice vectors that surrround the central one.
This is generally true along the Periodic Moir\'e Pattern line.
Indeed, the number of moir\'{e} reciprocal lattice vectors in $n$ consecutive hexagonal shells
that start from a central reciprocal lattice vector is $3 n (n+1)$,
which by (\ref{l_M}) is precisely $l-1$ when $m = n+1$.

Last, let us discuss the periodic boundary conditions, to be assumed hereafter.
The spatial periods shall be ${\bi L}_1 = N_1 {\bi A}_1$ and ${\bi L}_2 = N_2 {\bi A}_2$,
where $N_1$ and $N_2$ are positive integers.
Recall the identities ${\bi B}_i\cdot{\bi A}_{\bar j} = (-1)^{i-1} 2\pi \delta_{i,j}$,
where ${\bar 1} =2$ and ${\bar 2} = 1$.
An allowed wavenumber ${\bi k}$ must satisfy
$e^{i{\bi k}\cdot{\bi L_1}} = 1 = e^{i{\bi k}\cdot{\bi L_2}}$.
This yields that the allowed wave numbers are
\begin{equation}
{\bi k} = {n_1\over N_2} {\bi B}_1 + {n_2\over N_1} {\bi B}_2 ,
\label{allowed_wave}
\end{equation}
where $n_1$ and $n_2$ are any integers.

\subsection{Electron Hopping Hamiltonian\label{e_hppng_hmltnn}}
We shall model the electronic structure of twisted bi-layer graphene at commensurate twist angles by
electron hopping within ($\parallel$) and in between ($\perp$) the twisted sheets of graphene.
The nearest neighbors within each sheet of graphene are consecutive A and B sites on the honeycomb,
while nearest neighbors in between twisted sheets of graphene
are achieved by twisting A/B and A/A stacked bi-layer graphene.
Figures \ref{twist_A_B} and \ref{twist_A_A} show this, respectively.
The matrix element for electron hopping between nearest-neighbors in graphene is called $-t_{\parallel}$,
while the matrix element for a nearest-neighbor link in between twisted sheets of graphene is
 called $-t_{\perp}^{(0)}$ or $-t_{\perp}^{(1)}$.
Here, the superscript $0$ refers to A-A or B-B links, 
while the superscript $1$ refers to A-B or B-A links.

\begin{figure}
\includegraphics[scale=1.00, angle=0]{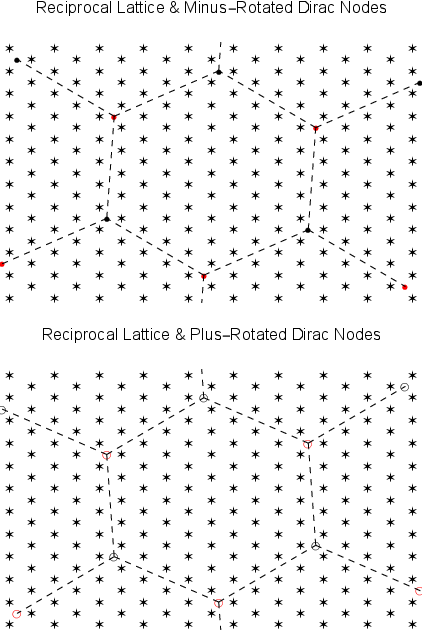}
\caption{Moir\'{e} reciprocal lattice ($*$) of bi-layer graphene
 at commensurate twist angle equal to $\theta = 7.3^{\circ}$, 
which corresponds to integer parameters $m=5$ and $n=4$.
Also shown are the Dirac nodes corresponding to 
the minus-rotated ($\cdot$) and plus-rotated ($\circ$) sheets of graphene. 
The dashed lines represent the edge of the first Brillouin zone of the rotated sheets of graphene.
Both Brillouin zones of the graphene sheets in isolation contain $l=61$ moir\'{e}
reciprocal lattice vectors.}
\label{rcprcl_lttc_b}
\end{figure}

The intra-graphene-sheet  contribution to the hamiltonian of an electron with crystal momentum ${\bi k}$ is 
$H_{++} ({\bi k}) + H_{--} ({\bi k})$,
where
\begin{equation}
H_{\pm\pm} ({\bi k}) = -t_{\parallel}
\sum_{i=0}^{l-1} [\zeta_{\pm}({\bi k} - {\bi G}_i)
|{\bi k} - {\bi G}_i,A,\pm\rangle\rangle \langle\langle {\bi k} - {\bi G}_i,B,\pm | + {\rm H.C.}] .
\label{h_parallel}
\end{equation}
Here, 
the quantum numbers $+$ and $-$ denote the $+\alpha$-rotated and $-\alpha$-rotated sheets of graphene,
 respectively,
while ${\bi G}_0, {\bi G}_1, ..., {\bi G}_{l-1}$ are the moir\'{e} reciprocal lattice vectors
that lie inside the first Brillouin zone of a twisted sheet of graphene in isolation.
Figure \ref{rcprcl_lttc_b} shows these in the case of $l=61$ A(B) sites per sheet of graphene
in a moir\'{e} unit cell.
The $l$ moir\'{e} reciprocal lattice vectors are common to both sheets of graphene!
The null reciprocal lattice vector is ${\bi G}_0$.
Above, also,
\begin{equation}
\zeta_{\pm} ({\bi k}) = 
e^{i {\bi k}\cdot{\bi c_1(\pm)}} + e^{i {\bi k}\cdot{\bi c_2(\pm)}} + e^{i {\bi k}\cdot{\bi c_3(\pm)}}
\label{zeta}
\end{equation}
is the standard phase factor of the A-to-B site matrix element in graphene.
The hamiltonian (\ref{h_parallel}) is expressed in terms of plane-wave states
\begin{equation}
|{\bi k}, A(B), \pm\rangle\rangle =
{1\over{N^{1/2}}} \sum_{{\bi r}_{A(B) \pm}} e^{i{\bi k}\cdot {\bi r}_{A(B) \pm}} 
|{\bi r}_{A(B) \pm},A(B),\pm\rangle ,
\label{pln_wv}
\end{equation}
where $N = l\cdot N_M$ 
is the total number of A(B) sites, ${\bi r}_{A(B) \pm}$,
in the $\pm \alpha$-rotated sheet of graphene,
with $N_M = N_1 N_2$  true moir\'{e} unit cells.
Above, $|{\bi r}_{A(B) \pm}, A(B), \pm\rangle$ is a site state in twisted bi-layer graphene.

In real space,
the inter-graphene-sheet  contribution to the hamiltonian of an electron
has the form $H_{-+} + H_{+-}$,
where  $H_{-+}$ has four components:
\begin{eqnarray}
H_{A-,A+} &=& -\sum_{\delta{\bi r}_{A/A}} \sum_{\bi R} t_{\perp}^{(0)}
|{\bi R}+\delta{\bi r}_{A/A}(-),A,-\rangle \langle {\bi R}+\delta{\bi r}_{A/A}(+),A,+| \nonumber\\
%
%
%
H_{A-,B+} &=& -\sum_{\delta{\bi r}_{B/A}} \sum_{\bi R} t_{\perp}^{(1)}
|{\bi R}+{\bi C}_3+\delta{\bi r}_{B/A}(-),A,-\rangle \langle {\bi R}+{\bi C}_3+\delta{\bi r}_{B/A}(+),B,+| \nonumber\\
%
%
%
H_{B-,A+} &=& -\sum_{\delta{\bi r}_{A/B}} \sum_{\bi R} t_{\perp}^{(1)}
|{\bi R}-{\bi C}_2+\delta{\bi r}_{A/B}(-),B,-\rangle \langle {\bi R}-{\bi C}_2+\delta{\bi r}_{A/B}(+),A,+| \nonumber\\
%
%
%
H_{B-,B+} &=& -\sum_{\delta{\bi r}_{B/B}} \sum_{\bi R} t_{\perp}^{(0)}
|{\bi R}+\delta{\bi r}_{B/B}(-),B,-\rangle \langle {\bi R}+\delta{\bi r}_{B/B}(+),B,+| .
\label{H-+}
\end{eqnarray}
Above, ${\bi R}$ denote the twist centers for A-A and B-B nearest-neighbor links in between sheets of graphene
per true moir\'{e} unit cell, which are shown by Fig. \ref{cmmnsrt_tblg_7}.
The displacements $\delta{\bi r}_{A/A}(+)$  and $\delta{\bi r}_{A/A}(-)$, for example,
denote the plus-rotated A site of the A-A link and the minus-rotated A site of the A-A link 
in comparison to the A/A twist center. (See Fig. \ref{twist_A_A}.)
And notice that $\delta{\bi r}_{B/B}(\pm) = -\delta{\bi r}_{A/A}(\pm)$.
Also, the displacements $\delta{\bi r}_{B/A}(+)$  and $\delta{\bi r}_{B/A}(-)$, for example,
denote the plus-rotated B site of the B-A link and the minus-rotated A site of the B-A link
in comparison to the B/A twist center. (See Fig. \ref{twist_A_B}.)
The displacement ${\bi C}_3$ of the latter twist center compared to the former twist center
  must be added in this case.
Also notice that $\delta{\bi r}_{A/B}(\pm) = \delta{\bi r}_{B/A}(\pm)$.
The displacement $-{\bi C}_2$ of this last twist center compared to the first  twist center
  must be added in this case.
Last, $H_{+-} ({\bi k})$ is the hermitian conjugate of (\ref{H-+}) $H_{-+} ({\bi k})$.

To determine inter-graphene-sheet matrix elements
 of the hopping hamiltonian of the electron
with crystal momentum ${\bi k}$, $H_{-+} ({\bi k})$,
we substitute the inversion formula
\begin{equation}
|{\bi r}_{A(B) \pm},A(B),\pm\rangle =
{1\over{N^{1/2}}} \sum_{\bi k} e^{-i{\bi k}\cdot {\bi r}_{A(B) \pm}}           
|{\bi k}, A(B), \pm\rangle\rangle 
\label{site_stt}
\end{equation}
for the plane-wave states (\ref{pln_wv}) into (\ref{H-+}).
This yields the respective matrix elements:
\begin{eqnarray}
\langle\langle {\bi k}_i,A,-|H|{\bi k}_j,A,+\rangle\rangle &=& -\sum_{\delta{\bi r}_{A/A}}{t_{\perp}^{(0)}\over l}
e^{i[-{\bi k}_i\cdot\delta{\bi r}_{A/A}(-) + {\bi k}_j\cdot\delta{\bi r}_{A/A}(+)]} \nonumber\\
%
%
%
\langle\langle {\bi k}_i,A,-|H|{\bi k}_j,B,+\rangle\rangle &=& -e^{i({\bi G}_i-{\bi G}_j)\cdot{\bi C}_3}
\sum_{\delta{\bi r}_{B/A}} {t_{\perp}^{(1)}\over l}
e^{i[-{\bi k}_i\cdot\delta{\bi r}_{B/A}(-) + {\bi k}_j\cdot\delta{\bi r}_{B/A}(+)]} \nonumber\\
%
%
%
\langle\langle {\bi k}_i,B,-|H|{\bi k}_j,A,+\rangle\rangle &=& -e^{-i({\bi G}_i-{\bi G}_j)\cdot{\bi C}_2}
\sum_{\delta{\bi r}_{A/B}} {t_{\perp}^{(1)}\over l}
e^{i[-{\bi k}_i\cdot\delta{\bi r}_{A/B}(-) + {\bi k}_j\cdot\delta{\bi r}_{A/B}(+)]} \nonumber\\
%
%
%
\langle\langle {\bi k}_i,B,-|H|{\bi k}_j,B,+\rangle\rangle &=& -\sum_{\delta{\bi r}_{B/B}}{t_{\perp}^{(0)}\over l}
e^{i[-{\bi k}_i\cdot\delta{\bi r}_{B/B}(-) + {\bi k}_j\cdot\delta{\bi r}_{B/B}(+)]} ,
\label{h-+}
\end{eqnarray}
where ${\bi k}_i = {\bi k}-{\bi G}_i$ and ${\bi k}_j = {\bi k}-{\bi G}_j$.
Notice that ${\bi k}_i - {\bi k}_j$ is a moir\'{e} reciprocal lattice vector,
${\bi G}_j - {\bi G}_i$.
Also, the matrix elements that correspond to $H_{+-} ({\bi k})$ are the hermitian conjugates of the above.
This can be seen explicitly by swapping the $-$ and $+$ twist-layer quantum numbers in (\ref{h-+}).

The phase factors appearing in the off-diagonal A/B and B/A matrix elements (\ref{h-+}) are easily calculated.
Let ${\bi G} = n_1 {\bi B}_1 + n_2 {\bi B}_2$ be a moir\'{e} reciprocal lattice vector, 
where $n_1$ and $n_2$ are integers.  A direct calculation using expressions 
(\ref{C_i}) and (\ref{B_i}) yields the phase factor
$e^{i{\bi G}\cdot{\bi C}_1} = (e^{-i 2\pi/3})^{n_1+n_2}$.
The differences 
${\bi C}_2 - {\bi C}_1 = {\bi A}_3$ and ${\bi C}_3 - {\bi C}_1 = -{\bi A}_2$
then yields that
$e^{i{\bi G}\cdot{\bi C}_2}$ and $e^{i{\bi G}\cdot{\bi C}_3}$ 
are both equal to the previous result for $e^{i{\bi G}\cdot{\bi C}_1}$.
\ref{ppndx_a} shows how a suitable gauge transformation removes the above phase factors
from the off-diagonal A/B and B/A matrix elements (\ref{h-+}),
while they reappear in the corresponding A/A and B/B matrix elements.
This becomes convenient in the chiral limit\cite{chrl_lmt_tblg_prl_19}, where the latter matrix elements vanish.

We thereby conclude the construction of the electron hopping hamiltonian in momentum space.
It has the form
\begin{equation}
H ({\bi k}) =
\left[\matrix{
H_{--} ({\bi k}) & H_{-+} ({\bi k})\cr
H_{+-} ({\bi k}) & H_{++} ({\bi k})}\right] .
\label{mtrx_frm}
\end{equation}
The intra-graphene-sheet parts (\ref{h_parallel}), $H_{--}$ and $H_{++}$,
are $2\times 2$-block diagonal $2l\times 2l$ matrices, while the inter-graphene-sheet parts (\ref{h-+}),
 $H_{-+}$ and $H_{+-}$, are dense $2l\times 2l$ matrices.

\section{Degenerate Perturbation Theory\label{dgnrt_prtrbtn}}
\begin{figure}
\includegraphics[scale=1.00, angle=0]{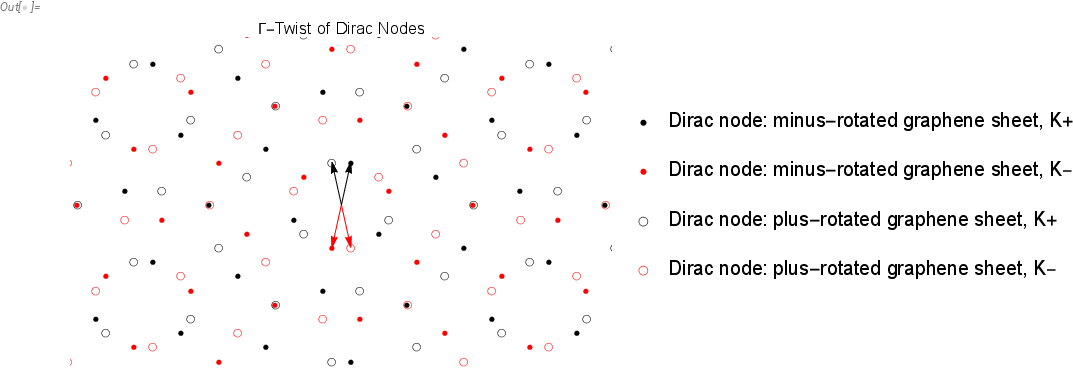}
\caption{Dirac nodes of graphene sheets that are twisted about the common $\Gamma$ point
in the first Brillouin zone by the largest commensurate angle $\theta = 21.8^{\circ}$.
It corresponds to integer parameters $m=2$ and $n=1$.
The black arrows and red arrows highlight the twist
at the $K+$ and $K-$ Dirac nodes of graphene, respectively.
Comparison with Fig. \ref{cmmnsrt_tblg_7} for the corresponding lattice in real space reveals
{\it duality}: rotation by $90^{\circ}$.}
\label{g_twst_7}
\end{figure}

Figure \ref{g_twst_7} shows the Dirac nodes of bi-layer graphene twisted 
by a commensurate angle equal to $\theta = 21.8^{\circ}$
about a common $\Gamma$ point in the first Brillouin zone.
The lattice of Dirac nodes in momentum space is equivalent to the lattice in real space of
twisted bi-layer graphene, Fig. \ref{cmmnsrt_tblg_7}, but rotated by $90^{\circ}$.
In the absence of hopping between sheets of graphene,
the lowest-energy states are states near the coincidence of Dirac nodes in each sheet,
black dot in red circle, or red dot in black circle,
which mix valley quantum numbers.
Below, we shall apply degenerate perturbation theory to such states,
and thereby reveal the latter.  Following Bistritzer and MacDonald\cite{bistritzer_macdonald_11},
low-energy degenerate states also exist at momenta
 that bisect the twisted $K$ points in the first Brilouin zone.
Below, we shall also apply degenerate perturbation theory
to such states, and thereby reveal the appearance of flat bands at a magic twist angle.

\subsection{Electron Energy Spectrum near Corner of Moir\'{e} Brillouin Zone}
The valley quantum numbers mix in twisted bi-layer graphene at commensurate twist angles.
Figure \ref{g_twst_7} shows the corresponding graphene nodes at
the largest commensurate twist angle of $\theta = 21.8^{\circ}$.
It is analogous to Fig. \ref{cmmnsrt_tblg_7} for twisted bi-layer graphene in real space,
but rotated by $90^{\circ}$.
Here, the $\Gamma$ point in the first Brillouin zone of graphene 
is the common twist center for the two sheets of graphene.
In general, assume bi-layer graphene with a commensurate twist angle
such that the difference of the integer parameters $m-n$ is not a multiple $3$.
This gaurantees the three types of twist centers evident in Fig. \ref{cmmnsrt_tblg_7}.
(See section \ref{cmmnsrt_tblg}.)
By analogy with such twisted bi-layer graphene in real space,
the momenta
$-m {\bi d}_2(-) + n {\bi d}_3(-)$ (red dot) and $-n {\bi d}_2(+) + m {\bi d}_3(+)$ (black circle)
coincide, for example, at a point along 
the $x$ axis that we shall name $+K_{\widebar{M}} {\hat{\bi x}}$ (red dot in black circle).
It follows the scaling law $K_{\widebar{M}} = \sqrt{l}\, K$, where $l$ is given by (\ref{l_M}).
An isolated sheet of graphene has two types of Dirac nodes in the energy spectrum:
Dirac nodes at $+K {\hat{\bi y}}$ up to a reciprocal lattice vector, which we label by $K+$, and
Dirac nodes at $-K {\hat{\bi y}}$ up to a reciprocal lattice vector, which we label by $K-$.
The $+\alpha$-rotated Dirac nodes at $K+$ are degenerate with
the $-\alpha$-rotated Dirac nodes at $K-$ at this point in momentum space, $K_{\widebar{M}}+$. 
Vice versa, the momenta
$m {\bi d}_2(-) - n {\bi d}_3(-)$ (black dot)
and $n {\bi d}_2(+) - m {\bi d}_3(+)$ (red circle) 
coincide, for example, at a point along the $x$ axis $-K_{\widebar{M}} {\hat{\bi x}}$ (black dot in red circle).
The $+\alpha$-rotated Dirac nodes at $K-$ are degenerate with 
the $-\alpha$-rotated Dirac nodes at $K+$ at this point in momentum space, $K_{\widebar{M}}-$. 

In the limit of weak hopping in between the sheets of graphene,
we can apply degenerate perturbation theory
at momenta near the point $+K_{\widebar{M}} {\hat{\bi x}}$ identified above.
(See Fig. \ref{g_twst_7}, red dot in black circle.)
Recall the Hamiltonian in momentum space constructed previously in section \ref{e_hppng_hmltnn}:
(\ref{mtrx_frm}) with matrix elements (\ref{h_parallel}) and (\ref{h-+}).
The twist about the $\Gamma$ point depicted by Fig. \ref{g_twst_7}
coincides with truncating the hopping hamiltonian to momenta
${\bi k}_i = {\bi k}-{\bi G} = {\bi k}_j$,
where ${\bi G}$ is a reciprocal lattice vector inside a graphene Brillouin zone.
This yields a truncated $4\times 4$ hamiltonian of the form
\begin{equation}
H_{K_M+} (\delta{\bi k}) =
\left[ {\begin{array}{cccc}
0     & z_-^*   & w_0^{\prime} & w_1^{\prime}   \\
z_-   & 0       & w_1^{\prime} & w_0^{\prime *} \\
w_0^{\prime *}  & w_1^{\prime *} & 0      & z_+   \\
w_1^{\prime *}    & w_0^{\prime} & z_+^*  & 0     \\
\end{array} } \right] .
\label{H_KM+}
\end{equation}
Here, we express the momentum as
${\bi k} - {\bi G} = K_{\widebar{M}} {\hat{\bi x}} + \delta {\bi k}$.
Above, we have intra-graphene matrix elements
 $z_{\pm} = - t_{\parallel} \zeta_{\pm}({\bi k}-{\bi G})$
 at momenta ${\bi k}-{\bi G}$ in the vicinity of $K_{\widebar{M}} {\hat{\bi x}}$.
It is convenient to re-write the phase factor (\ref{zeta}) as
$\zeta_{\pm}({\bi k}) = e^{i{\bi k}\cdot{\bi c}_1(\pm)} \zeta_{\pm}^{\prime} ({\bi k})$, where
$\zeta_{\pm}^{\prime} ({\bi k}) =
1 + e^{i {\bi k}\cdot[{\bi c}_2(\pm)-{\bi c}_1(\pm)]} + e^{i {\bi k}\cdot[{\bi c}_3(\pm)-{\bi c}_1(\pm)]}$,
or equivalently
by the identities 
${\bi a}_1 = {\bi c}_2 - {\bi c}_3$, 
${\bi a}_2 = {\bi c}_3 - {\bi c}_1$, and
${\bi a}_3 = {\bi c}_1 - {\bi c}_2$,
\begin{equation}
\zeta_{\pm}^{\prime} ({\bi k}) = 
1 + e^{-i{\bi k}\cdot{\bi a}_3(\pm)} + e^{i {\bi k}\cdot{\bi a}_2(\pm)} .
\label{zeta_prime}
\end{equation}
Because expression (\ref{zeta_prime}) is manifestly periodic over 
the reciprocal lattice of the $\pm$-rotated sheet of graphene,
Fig. \ref{g_twst_7} indicates the equalities
$\zeta_{+}^{\prime} ({\bi k}-{\bi G}) = \zeta_{+}^{\prime} ({\bi d}_1(+)+\delta {\bi k})$
and
$\zeta_{-}^{\prime} ({\bi k}-{\bi G}) = \zeta_{-}^{\prime} (-{\bi d}_1(-)+\delta {\bi k})$,
Hence, we have the respective Dirac cones
$\zeta_{+}^{\prime} ({\bi d}_1(+)+\delta {\bi k}) \cong i {3\over 2} a (\delta k_x + i\delta k_y) e^{-i\alpha}$
and
$\zeta_{-}^{\prime} (-{\bi d}_1(-)+\delta {\bi k}) \cong i {3\over 2} a (\delta k_x - i\delta k_y) e^{-i\alpha}$.
Second, it is shown in \ref{ppndx_b} that the phase factor 
at the degeneracy momentum $K_{\widebar{M}} {\hat{\bi x}}$ is a cube-root of unity:
$e^{+i K_{\widebar{M}} {\hat{\bi x}}\cdot {\bi c}_1(\pm)} = {\rm exp}[i 2\pi (m+n)/3]$.
We thereby obtain the following expressions for
the intra-graphene matrix elements of the truncated hamiltonian (\ref{H_KM+}):
\begin{eqnarray}
z_- = - v_D e^{i(\alpha-\delta)} i^* (\delta k_x + i\delta k_y) , \nonumber \\
z_+ = - v_D e^{-i(\alpha-\delta)} i (\delta k_x + i\delta k_y) .
\label{z_+K_M}
\end{eqnarray}
Here, $v_D = {3\over 2} a\, t_{\parallel}$ is the velocity of a Dirac cone,
and $\delta = 2\pi (m+n)/3$ is a $3$-fold symmetric phase angle.
Along the Periodic Moir\'e Pattern line, $m = n+1$, the latter is equal to $\delta = 2\pi (2 n + 1)/3$,
which implies
$e^{i\delta} = e^{-i 2\pi (n-1)/3}$.

Next, we shall compute the inter-graphene-sheet matrix element  $w_1^{\prime}$ of 
the truncated hamiltonian (\ref{H_KM+}).
It is given by the expression
\begin{eqnarray}
\langle\langle{\bi k}_i,A,-|H|{\bi k}_i,B,+\rangle\rangle
= - \sum_{\delta{\bi r}_{B/A}} {t_{\perp}^{(1)} \over l} 
e^{- i {\bi k}_i\cdot[\delta{\bi r}_{B/A}(-)-\delta{\bi r}_{B/A}(+)]} \nonumber \\
\label{h_perp_AB}
\end{eqnarray}
for inter-graphene-sheet matrix elements (\ref{h-+}) 
at momenta ${\bi k}_i = {\bi k}-{\bi G}$
near $K_{\widebar{M}} {\hat{\bi x}}$.
Henceforth, we shall confine the integer parameters to the Periodic Moir\'e Pattern line,
$m = n + 1$,
in which case the A-B links form $n$ consecutive hexagonal shells within the Wigner-Seitz unit cell
centered at the A-B Bernal stack. (See Fig. \ref{twist_A_B}.)
Using $\delta{\bi r}_{B/A}(\pm) = n_1 {\bi a}_1(\pm) + n_2 {\bi a}_2(\pm)$ ,
where $n_1$ and $n_2$ are integers,
we have that the arguments of the oscillatory exponentials in (\ref{h_perp_AB}) 
have the form imaginary $i$ times
$(K_{\widebar{M}} {\hat{\bi x}}\cdot\delta{\bi a}_1) n_1 + (K_{\widebar{M}} {\hat{\bi x}}\cdot\delta{\bi a}_{2}) n_2$,
where
\begin{eqnarray}
\delta{\bi a}_1 &= {\bi a}_1(+) - {\bi a}_1(-) &= - a_{\triangle} (2 \sin \alpha) {\hat{\bi x}} , \nonumber \\
\delta{\bi a}_2 &= {\bi a}_2(+) - {\bi a}_2(-) &= a_{\triangle} (2 \sin \alpha) 
\Bigg({1\over 2} {\hat{\bi x}} - {\sqrt{3}\over 2} {\hat{\bi y}}\Biggr) , \nonumber \\
\delta{\bi a}_3 &= {\bi a}_3(+) - {\bi a}_3(-) &= a_{\triangle} (2 \sin \alpha)
\Biggl({1\over 2} {\hat{\bi x}} + {\sqrt{3}\over 2} {\hat{\bi y}}\Biggr) .
\label{da_i}
\end{eqnarray}
Recall, now, that $K_{\widebar{M}} = \sqrt{l}\, K$ and $K = 4\pi/3 a_{\triangle}$.
Three-fold symmetric terms are grouped together by adding such oscillatory exponential terms
with arguments proportional to
$n_1 \delta{\bi a}_1 + n_2 \delta{\bi a}_2$, 
$n_1 \delta{\bi a}_2 + n_2 \delta{\bi a}_3$,
and $n_1 \delta{\bi a}_3 + n_2\delta{\bi a}_1$.
After applying $6$-fold rotation symmetry of the right-hand side of (\ref{h_perp_AB}),
and using the following results for the dot products,
\begin{eqnarray}
K_{\widebar{M}} {\hat{\bi x}}\cdot\delta{\bi a}_1 &= - 4 \pi / 3 , \nonumber \\
K_{\widebar{M}} {\hat{\bi x}}\cdot\delta{\bi a}_2 &= + 2 \pi /3  , \nonumber \\
K_{\widebar{M}} {\hat{\bi x}}\cdot\delta{\bi a}_3 &= + 2 \pi /3  ,
\label{dot_products_K_M_a}
\end{eqnarray}
we obtain the result
\begin{equation}
w_1^{\prime} = -{t_{\perp}^{(1)}\over l}\Biggl[1
+ 6\, {\rm Re} \sum_{n_1 = 1}^n \sum_{n_2=0}^{n_1 - 1} (e^{i 2\pi/3})^{n_1+n_2}\Biggr] .
\label{w_1_prime}
\end{equation}
for (\ref{h_perp_AB}) at $+K_{\widebar{M}} {\hat{\bi x}}$
in the simple case where $t_{\perp}^{(1)}$ is a constant over the Wigner-Seitz unit cell.
Above, in (\ref{dot_products_K_M_a}), we have used the result $\sqrt{l} (2 \sin \alpha) = 1$ at  $m = n + 1$,
by (\ref{sine}) and (\ref{l_M}).
Because the A-B twist centers  show $6$-fold rotation symmetry,
which includes inversion symmetry (see Fig. \ref{cmmnsrt_tblg_7}),
the matrix element $w_1^{\prime}$ is necessarily real.
Importantly, notice that the oscillatory exponential factor above in general
will result in a small matrix element of order $t_{\perp}^{(1)} / l$.  
A direct summation of the geometric series (\ref{w_1_prime})  yields the particular result
$w_1^{\prime} = 2 (t_{\perp}^{(1)} / l) [\cos 2\pi (n-1)/3]$.
(See \ref{ppndx_c}.)
The small inter-valley matrix element $w_1^{\prime}$ 
is consistent with the Bistritzer-MacDonald method\cite{bistritzer_macdonald_11},
in which case $w_1^{\prime}$ is neglected.

Next, we shall compute the inter-graphene-sheet matrix element  $w_0^{\prime}$ of 
the truncated hamiltonian (\ref{H_KM+}).
It is given by the expression
\begin{eqnarray}
\langle\langle{\bi k}_i,A,-|H|{\bi k}_i,A,+\rangle\rangle
= - \sum_{\delta{\bi r}_{A/A}} {t_{\perp}^{(0)} \over l} 
e^{- i {\bi k}_i\cdot[\delta{\bi r}_{A/A}(-)-\delta{\bi r}_{A/A}(+)]} \nonumber \\
\label{h_perp_AA}
\end{eqnarray}
for inter-graphene-sheet matrix elements (\ref{h-+}) 
at momenta ${\bi k}_i = {\bi k}-{\bi G}$
near $K_{\widebar{M}} {\hat{\bi x}}$.
The displacements of the links in between the sheets of graphene from the center of the Wigner-Seitz unit cell
are given by $\delta{\bi r}_{A/A}(\pm) = m_1 {\bi c}_1(\pm) + m_2 {\bi c}_2(\pm)$ ,
where  $m_1$ and $m_2$ are integers that satisfy the condition $m_1+m_2\ {\rm mod}\ 3 = 1$.
Figure \ref{twist_A_A} displays that A-A links show  $3$-fold rotation symmetry.
This results in a null matrix element (\ref{h_perp_AA}) as we will now demonstrate.
Each group of three terms in the matrix element (\ref{h_perp_AA})
related by $120^{\circ}$ rotations is proportional to $\zeta_{A/A}^{(m_1,m_2)} (K_{\widebar{M}} {\hat{\bi x}})$, where
\begin{equation}
\zeta_{A/A}^{(m_1,m_2)} ({\bi k}) = e^{i{\bi k}\cdot\delta{\bi c}_0(m_1,m_2)} + 
e^{i{\bi k}\cdot\delta{\bi c}_+(m_1,m_2)} +
e^{i{\bi k}\cdot\delta{\bi c}_-(m_1,m_2)} ,
\label{zeta_AA}
\end{equation}
with
\begin{eqnarray}
\delta{\bi c}_0 (m_1, m_2) &= m_1 \delta{\bi c}_1 + m_2 \delta{\bi c}_2 , \nonumber \\
\delta{\bi c}_+ (m_1, m_2) &= m_1 \delta{\bi c}_2 + m_2 \delta{\bi c}_3 , \nonumber \\
\delta{\bi c}_- (m_1, m_2) &= m_1 \delta{\bi c}_3 + m_2 \delta{\bi c}_1 ,
\label{3_fold_c}
\end{eqnarray}
and where
\begin{eqnarray}
\delta{\bi c}_1 &= {\bi c}_1(+) - {\bi c}_1(-) &= a (2 \sin \alpha) {\hat{\bi y}} , \nonumber \\
\delta{\bi c}_2 &= {\bi c}_2(+) - {\bi c}_2(-) &= a (2 \sin \alpha)
\Biggl(-{\sqrt{3}\over 2}{\hat{\bi x}} - {1\over 2}{\hat{\bi y}}\Biggr) , \nonumber \\
\delta{\bi c}_3 &= {\bi c}_3(+) - {\bi c}_3(-) &= a (2 \sin \alpha) 
\Biggl({\sqrt{3}\over 2}{\hat{\bi x}} - {1\over 2}{\hat{\bi y}}\Biggr) .
\label{dc_i}
\end{eqnarray}
Using the following results for the dot products at $m - n = 1$,
\begin{eqnarray}
K_{\widebar{M}} {\hat{\bi x}}\cdot\delta{\bi c}_1 &= 0 , \nonumber \\
K_{\widebar{M}} {\hat{\bi x}}\cdot\delta{\bi c}_2 &= - 2 \pi /3  , \nonumber \\
K_{\widebar{M}} {\hat{\bi x}}\cdot\delta{\bi c}_3 &= + 2 \pi /3  ,
\label{dot_products_K_M_c}
\end{eqnarray}
direct calculation then yields
\begin{eqnarray}
\zeta_{A/A}^{(m_1, m_2)} (K_{\widebar{M}} {\hat{\bi x}}) &=& (e^{i 2\pi /3})^{m_1} + (e^{i 2\pi /3})^{m_2-m_1} + (e^{i 2\pi /3})^{-m_2} \nonumber \\
&=& (e^{i 2\pi /3})^{m_2-m_1}[(e^{i 2\pi /3})^{2 m_1 - m_2} + 1 + (e^{i 2\pi /3})^{m_1 - 2 m_2}] \nonumber \\
&=& (e^{i 2\pi /3})^{m_2-m_1}[(e^{i 2\pi /3})^{-(m_1 + m_2)} + 1 + (e^{i 2\pi /3})^{(m_1 +  m_2)}] . \nonumber \\
\label{zeta_perp_k_m}
\end{eqnarray}
It is equal to zero in the present case, where $(m_1 + m_2)$ mod $3 = 1$.
Similar three-fold rotation-symmetric calculation yields that 
the first non-trivial order of this matrix element with $\delta {\bi k}$ has the form
$w_0^{\prime} \cong  w_0^{\prime(1)} (\delta k_x - i\delta k_y)$.
In the simple case where $t_{\perp}^{(0)}$ is constant over the Wigner-Seitz unit cell,
direct summation of the resulting geometric series at $m - n = 1$ yields
${\rm lim}_{n\rightarrow\infty}  w_0^{\prime(1)} = -(a\, t_{\perp}^{(0)} / l) \sin [2\pi (n-1) /3]$.
(See \ref{ppndx_c}.)
Notice, importantly, that this group velocity of the matrix element is small, 
and of order $a t_{\perp}^{(0)} / l$, or zero.
It is also thereby consistent with the Bistritzer-MacDonald method\cite{bistritzer_macdonald_11},
which neglects such transitions.

\begin{figure}
\includegraphics[scale=1.00, angle=0]{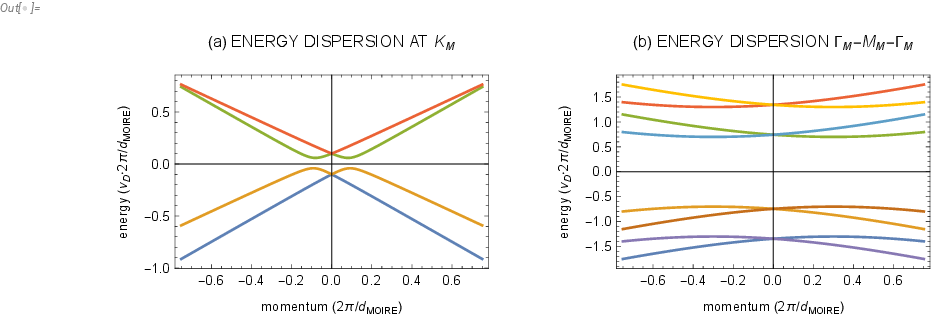}
\caption{(a) Dispersion about momentum $K_M$ of low-energy bands
as predicted by degenerate perturbation theory, (\ref{chrctrstc_eq}) and (\ref{a'_n}).
Here, we have set $\sin (\alpha-\delta) = \sqrt{3} /2$,
and $w_0^{\prime (1)} = -0.1\, v_D$ and $w_1^{\prime} = -0.1\, v_D \cdot 2\pi/d_{\rm moire}$,
with $d_{\rm moire} = \sqrt{l} d$, where $d = {\sqrt{3}\over 2} a_{\triangle}$.
(b) Dispersion along $\Gamma_M - M_M - \Gamma_M$ of low-energy bands
as predicted by degenerate perturbation theory, (\ref{chrctrstc_eq}) and (\ref{a_n}).
Here, we have set $w_0^{(0)} = -0.3\, v_D \cdot 2\pi/d_{\rm moire} = w_1$.}
\label{dgnrt_prtrbtn_K&M}
\end{figure}

We shall now show that inter-valley mixing results in a small band gap for
twisted bi-layer graphene at commensurate twist angles along the Periodic Moir\'e Pattern line.
The characteristic equation associated with the truncated $4\times 4$ hamiltonian (\ref{H_KM+}) is
%
\begin{equation}
0 = \varepsilon^4 + a_2 \varepsilon^2 + a_1 \varepsilon + a_0 ,
\label{chrctrstc_eq}
\end{equation}
with coefficients
\begin{eqnarray}
a_2 &=& - 2 ( |w_0^{\prime}|^2 + w_1^{\prime 2} + |z|^2 ) , \nonumber \\
a_1 &=& - 2 [w_0^{\prime} (z_+ + z_-) + w_0^{\prime *} (z_+^* + z_-^*)] w_1^{\prime} , \nonumber \\
a_0 &=& (w_1^{\prime 2} - |w_0^{\prime}|^2)^2 - w_1^{\prime 2} (z_+ z_-^* + z_+^* z_-) \nonumber \\
&& - w_0^{\prime 2} z_+ z_- - w_0^{\prime * 2} z_+^* z_-^*  +  |z|^4 .
\label{cffcnts_K_M}
\end{eqnarray}
Above, $w_1^{\prime}$ is assumed to be real, and $|z_+| = |z| = |z_-|$.
Plugging in expressions (\ref{z_+K_M}) and $w_0^{\prime} = w_0^{\prime (1)} (\delta k_x - i \delta k_y)$
yields the following expressions for the above coefficients:
\begin{eqnarray}
a_2 &=& - 2[w_1^{\prime 2} + (v_D^2 + w_0^{\prime (1) 2}) |\delta{\bi k}|^2] , \nonumber \\
a_1 &=& 8 [\sin (\alpha-\delta)]w_1^{\prime} w_0^{\prime (1)} v_D  |\delta{\bi k}|^2 , \nonumber \\
a_0 &=&  w_1^{\prime 4} +
 2 w_1^{\prime 2} \{[\cos 2(\alpha-\delta)] v_D^2 - w_0^{\prime (1) 2}\} |\delta{\bi k}|^2 + 
 (v_D^2 - w_0^{\prime (1) 2})^2 |\delta{\bi k}|^4 .
\label{a'_n}
\end{eqnarray}
%
At $|w_0^{\prime (1)}| \ll v_D$,
the first-order term $a_1 \varepsilon$ in (\ref{chrctrstc_eq}) can be neglected at large enough momentum $|\delta{\bi k}|$.
This yields the dispersions in energy:
\begin{equation}
\varepsilon_{\pm}^{(0) 2} = (v_D |\delta {\bi k}| \pm [\sin(\alpha-\delta)] w_1^{\prime})^2 
+ ([\cos(\alpha-\delta)] w_1^{\prime})^2 .
\label{epsilon_0^2}
\end{equation}
It reveals that such twisted bi-layer graphene is an insulator at charge neutrality,
 with a small gap in the energy bands:
$2 \Delta_0 = 2 |[\cos(\alpha-\delta)] w_1^{\prime}|$.
Notice that the above approximation for the low-energy spectrum (\ref{epsilon_0^2}) shows particle-hole symmetry.
This  is an artifact.  Treating the former first-order term in the characteristic equation (\ref{chrctrstc_eq})
as a perturbation results in a first-order correction (\ref{eps_1}) to $\varepsilon_{\pm}^{(0)}$
that breaks particle-hole symmetry.
Figure \ref{dgnrt_prtrbtn_K&M} (a) shows the energy dispersion around $K_M$
that is predicted by degenerate perturbation theory
upon a numerical solution of (\ref{chrctrstc_eq}) and (\ref{a'_n}).
As expected, particle-hole symmetry is broken.

Last, what is the nature of inter-valley mixing at the opposing point in momentum,
$K_{\widebar{M}} {\hat{\bi x}}\rightarrow -K_{\widebar{M}} {\hat{\bi x}}$ ?
It can easily be shown that the graphene matrix elements in the corresponding
truncated $4\times 4$ hamiltonian (\ref{z_+K_M}) are negative complex conjugates of each other:
$z_{\pm} \rightarrow - z_{\pm}^*$.
Furthermore, the inter-graphene sheet matrix elements (\ref{h-+}) 
are given by $w_0^{\prime *}(-{\bi k})$ and $w_1^{\prime *}(-{\bi k})$.
Hence, we have the replacements in $w_0^{\prime}$ :
$\delta k_x + i \delta k_y \rightarrow \delta k_x - i \delta  k_y$ 
 and $w_0^{\prime (1)} \rightarrow - w_0^{\prime (1)}$.
The corresponding A/B inter-sheet matrix element $w_1^{\prime}$ is a non-zero real number
at $K_{\widebar{M}} {\hat{\bi x}}$, and it therefore remains unchanged.
Inspection of the characteristic equation (\ref{chrctrstc_eq}) and (\ref{cffcnts_K_M}) 
plus the above replacements yields that the energy eigenvalues remain unchanged:
$\varepsilon \rightarrow \varepsilon$.

\begin{figure}
\includegraphics[scale=1.00, angle=0]{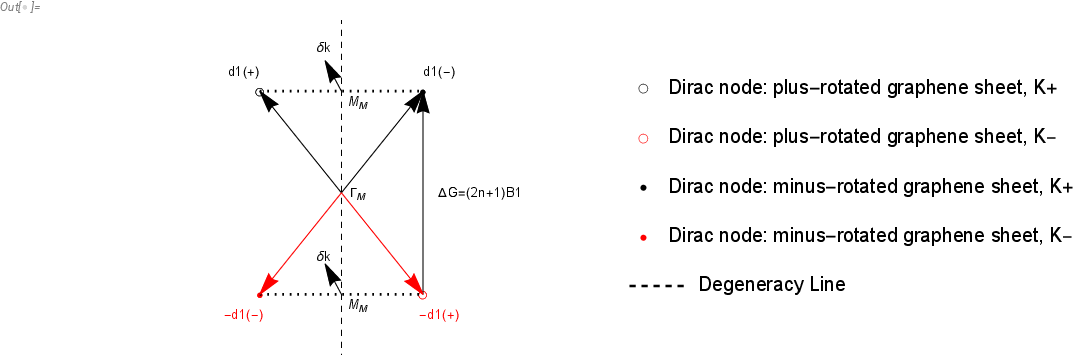}
\caption{Schematic depiction of ``twisted'' Dirac nodes 
following Bistritzer and MacDonald ref. \cite{bistritzer_macdonald_11}.
They are bisected by the (degeneracy) line $M_M$-$\Gamma_M$-$M_M$ in momentum space.}
\label{dgnrt_prtrbtn_M_M}
\end{figure}

\subsection{Electron Energy Spectrum near Middle of Edge of Moir\'e Brillouin Zone\label{dgnrt_prtrbtn_thry_M_M}}
We shall now reveal the origin of the band-flattening effect in twisted bi-layer graphene
by analyzing the hopping hamiltonian at momenta in 
the vicinity of $M_M$ within degenerate perturbation theory.
We shall, in particular, follow the idea proposed
 by Bistritzer and MacDonald \cite{bistritzer_macdonald_11} of ``twisting'' the Dirac nodes.
The black arrows in Fig. \ref{g_twst_7} correspond to these (\ref{d1}):
${\bi d}_1(+)$ and ${\bi d}_1(-)$.
Because the $k_y$ momentum axis bisects the latter two Dirac nodes,
any momentum point on that axis has corresponding graphene energy eigenvalues that are degenerate.
Here, the origin in momentum space is defined by the twist center
of the momentum arrows in Fig. \ref{g_twst_7}.
The same is true of the pair of Dirac nodes at $-{\bi d}_1(+)$ and $-{\bi d}_1(-)$,
which are the red arrows in Fig. \ref{g_twst_7}.
Figure \ref{dgnrt_prtrbtn_M_M} summarizes the previous.
Below, we will work out the degenerate perturbation theory of these two pairs of Dirac nodes.

We will now apply degenerate perturbation theory
in the limit of weak hopping in between the sheets of graphene,
at momenta ${\bi k} - {\bi G}$ near the point $+M_M {\hat{\bi y}}$
that bisects the ``twisted'' Dirac nodes located at ${\bi d}_1 (+)$ and ${\bi d}_1 (-)$.
Here, $M_M = (\cos \alpha) K$.
This approximation is described by 
the truncated $4\times 4$ hamiltonian of the form
\begin{equation}
H_{M_M+} (\delta{\bi k}) =
\left[ {\begin{array}{cccc}
0     & z_-   & w_0 & w_1   \\
z_-^*   & 0       & w_1 & w_0^* \\
w_0^*  & w_1^* & 0      & z_+   \\
w_1^*    & w_0 & z_+^*  & 0     \\
\end{array} } \right] .
\label{H_MM+}
\end{equation}
Here, we express the momentum as
${\bi k} - {\bi G} = M_M {\hat{\bi y}} + \delta {\bi k}$.
Above, we have intra-graphene matrix elements
 $z_{\pm} = - t_{\parallel} \zeta_{\pm}({\bi k}-{\bi G})$
 at momenta ${\bi k}-{\bi G}$ in the vicinity of $M_M {\hat{\bi y}}$.
As in the previous case, we approximate
$\zeta_{\pm}({\bi k}-{\bi G}) = 
e^{i{\bi d}_1(\pm) \cdot {\bi c}_1(\pm)} \zeta_{\pm}^{\prime}({\bi d}_1(\pm)+\delta{\bi k})$
at the Dirac nodes.
Notice, first, that the phase factor is unity
because ${\bi d}_1(\pm)$ and ${\bi c}_1(\pm)$ are perpendicular.
Also, $\zeta_{\pm}^{\prime} ({\bi k}-{\bi G})$ 
is periodic over the $\pm$-rotated reciprocal lattice by (\ref{zeta_prime}).
We therefore have the intra-graphene matrix elements
\begin{eqnarray}
z_- = - v_D e^{i\alpha} i (k_{-,x} + i k_{-,y}) , \nonumber \\
z_+ = - v_D e^{-i\alpha} i (k_{+,x} + i k_{+,y}) .
\label{z_+M_M}
\end{eqnarray}
Above, ${\bi k}_{\pm} = {\bi k}-{\bi G}-{\bi d}_1(\pm)$.

We shall next compute the inter-graphene A-B matrix element 
$w_1$ in the truncated $4\times 4$ hamiltonian (\ref{H_MM+}).
This is achieved by repeating the previous calculation of $w_1^{\prime}$ at momentum $K_{\widebar{M}} {\hat{\bi x}}$
(\ref{h_perp_AB})-(\ref{dot_products_K_M_a}),
but instead  at momentum $M_M {\hat{\bi y}}$.
The relevant dot products in this case are:
\begin{eqnarray}
M_M {\hat{\bi y}}\cdot\delta{\bi a}_1 &= 0 , \nonumber \\
M_M {\hat{\bi y}}\cdot\delta{\bi a}_2 &= - (2 \pi /3) \sqrt{3} (\sin \theta)  , \nonumber \\
M_M {\hat{\bi y}}\cdot\delta{\bi a}_3 &= + (2 \pi /3) \sqrt{3} (\sin \theta) .
\label{dot_products_M_M_a}
\end{eqnarray}
After applying $3$-fold rotation symmetry to 
the sum over hopping links in between sheets of graphene (\ref{h_perp_AB}),
we obtain the result
\begin{equation}
w_1 = -{t_{\perp}^{(1)}\over l}\Biggl[1
+ 2\, {\rm Re} \sum_{n_1 = 1}^n \sum_{n_2=0}^{n_1 - 1}
(z_0^{-n_2} + z_0^{n_2-n_1} + z_0^{n_1})\Biggr]
\label{w_1}
\end{equation}
in the simple case that $t_{\perp}^{(1)}$ is uniform,
where $z_0 = {\rm exp} [i (2 \pi /3) \sqrt{3} (\sin \theta)]$.
Again, A-B twist centers also show space-inversion symmetry,
which implies that $w_1$ must be real.
Along the Periodic Moir\'e Pattern line, $m=n+1$, it is easily shown from (\ref{sine}) that
${\rm lim}_{n\rightarrow\infty} z_0 = {\rm exp} \{ i 2 \pi / [3 (n+{1\over 2})] \}$.
Direct evaluation of the geometric series sums (\ref{w_1}) in this limit yields the result
\begin{equation}
{\rm lim}_{n\rightarrow\infty} w_1 =
- {3\over{2\pi}}\Biggl({1\over{\sqrt{3}}} + {3\over{2\pi}}\Biggr) t_{\perp}^{(1)} 
\cong - 0.5036\, t_{\perp}^{(1)} .
\label{w_1_lim}
\end{equation}
%
This is shown in \ref{ppndx_c_w0_w1}
from closed-form expressions of the geometric series sums (\ref{w_1}).

\begin{figure}
\includegraphics[scale=1.00, angle=0]{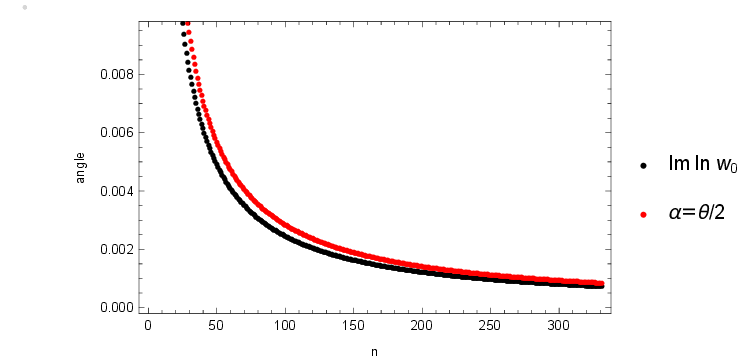}
\caption{Comparison of ${\rm Im}\, ln\, w_0$
 and half the twist angle, $\alpha$,
versus $n$ along the Periodic Moir\'e Pattern line, $m = n+1$.
The nearest-neighbor hopping matrix elements $t_{\perp}^{(0)}$ between the two graphene sheets are uniform.
See \ref{ppndx_c_w0_w1} for closed form expressions of $w_0$.}
\label{angle_vs_n}
\end{figure}

Next, we shall compute the matrix element between A sites of different sheets of graphene,
$w_0$, in the truncated $4\times 4$ hamiltonian (\ref{H_MM+}).
As in the previous subsection,
it is given by the expression (\ref{h_perp_AA}),
but evaluated at momenta ${\bi k}_i = {\bi k} - {\bi G}$ near $M_M {\hat{\bi y}}$ instead.
Recall that $M_M = (\cos \alpha) K$.
Again, it is useful to compute the matrix element in a $3$-fold symmetric way,
as described by expressions (\ref{zeta_AA})-(\ref{dc_i}),
where $m_1$ and $m_2$ are integers that satisfy the condition $(m_1 + m_2)$ mod $3 = 1$.
Using the following results for the dot products,
\begin{eqnarray}
M_M {\hat{\bi y}}\cdot\delta{\bi c}_1 &= {4\pi\over 3} {\sin \theta\over{\sqrt{3}}} , \nonumber \\
M_M {\hat{\bi y}}\cdot\delta{\bi c}_2 &= -{2\pi\over 3} {\sin \theta\over{\sqrt{3}}} , \nonumber \\
M_M {\hat{\bi y}}\cdot\delta{\bi c}_3 &= -{2\pi\over 3} {\sin \theta\over{\sqrt{3}}} ,
\label{dot_products_M_M_c}
\end{eqnarray}
direct calculation then yields
\begin{equation}
\zeta_{A/A}^{(m_1, m_2)} (M_M {\hat{\bi y}}) 
= (z_0^{1/3})^{2 m_1 - m_2} + (z_0^{1/3})^{-m_1-m_2} + (z_0^{1/3})^{2 m_2-m_1} .
\label{zeta_perp_m_m}
\end{equation}
In the simple case where $t_{\perp}^{(0)}$ is uniform over the Wigner-Seitz unit cell,
numerical evaluation of the resulting closed-form expressions for the geometric series sums yields the form
$w_0 =  w_0^{(0)} e^{i \alpha}$  for the matrix element at momentum $+M_M {\hat{\bi y}}$,
where $w_0^{(0)}$ is real at large integer parameter $n$.
This  is demonstrated by Fig. \ref{angle_vs_n}.
It is also true that
the limit ${\rm lim}_{n\rightarrow\infty} w_0$ at momentum $+M_M {\hat{\bi y}}$
coincides with the corresponding limit (\ref{w_1_lim}) for $w_1$,
but with the replacement $t_{\perp}^{(1)} \rightarrow t_{\perp}^{(0)}$.
The above results are based on summations of geometric series
 in closed form that emerge from the summation
of (\ref{zeta_perp_m_m}) over the A-A links shown by Fig. \ref{twist_A_A}.
Those calculations are carried out in \ref{ppndx_c_w0_w1}.  

We shall now show how energy-level repulsion at the point $M_M$ on the edge of the first moir\'e Brillouin zone
can result in flat energy bands near zero energy.
The characteristic equation associated with the truncated $4\times 4$ hamiltonian (\ref{H_MM+}) 
again has the fourth-order polynomial form (\ref{chrctrstc_eq}),
but with coefficients
\begin{eqnarray}
a_2 &=& - 2 [|w_0|^2 + w_1^2 + (|z_-|^2  + |z_+|^2)/2] , \nonumber \\
a_1 &=& - 2 [w_0 (z_+ + z_-^*) + w_0^* (z_+^* + z_-)] w_1 , \nonumber \\
a_0 &=& (w_1^2 - |w_0|^2)^2 - w_1^2 (z_+ z_- + z_+^* z_-^*) \nonumber \\
&& - w_0^2 z_+ z_-^* - w_0^{* 2} z_+^* z_-  + |z_+|^2 |z_-|^2 .
\label{cffcnts_M_M}
\end{eqnarray}
Above, $w_1$ is assumed to be real.
The wavenumbers in the forms (\ref{z_+M_M}) for the intra-graphene matrix elements $z_+$ and $z_-$
follow   $k_{+,x} = + k_x$, $k_{-,x} = -k_x$, $k_{+,y} = \delta k_y = k_{-,y}$
along the $y$ axis that bisects the two Dirac nodes at ${\bi d}_1(+)$ and ${\bi d}_1(-)$.
(See Fig. \ref{g_twst_7}.)
Here, $k_x$ is the displacement of the $-$ Dirac node from the bisecting $y$ axis.
(See Fig. \ref{dgnrt_prtrbtn_M_M}.)
Substitution into  (\ref{cffcnts_M_M}) yields the corresponding coefficients
for the characteristic equation (\ref{chrctrstc_eq}):
\begin{eqnarray}
a_2 &=& - 2 [ w_0^{(0) 2} +  w_1^2 +  v_D^2 k_x^2 + v_D^2 (\delta k_y)^2] , \nonumber \\
a_1 &=& - 8 w_0^{(0)} w_1 v_D (\delta k_y) , \nonumber \\
a_0 &=& (w_1^2 - w_0^{(0) 2})^2 - 2 (w_1^2 - w_0^{(0) 2}) v_D^2 k_x^2 \nonumber \\
&& - 2 (w_1^2 + w_0^{(0) 2}) v_D^2 (\delta k_y)^2 + v_D^4 [k_x^2 + (\delta k_y)^2]^2 .
\label{a_n}
\end{eqnarray}
%
At the point $M_M$ in momentum space, we have $\delta k_y = 0$.
This yields the coefficients $a_1 = 0$
and $a_0 = [w_1^2 - w_0^{(0) 2} - (v_D k_x)^2]^2$.
The energy eigenvalues are then easily obtained, and they are given by
\begin{equation}
\varepsilon_{\pm}^{(0) 2}(M_M) = -{a_2\over 2}\pm\sqrt{\Biggl({a_2\over 2}\Biggr)^2 - a_0} .
\label{eps_0}
\end{equation}
Above, $\varepsilon_-^{(0)}(M_M)$ and $\varepsilon_+^{(0)}(M_M)$ 
are closer and farther away from zero energy, respectively.
Because $a_1 = 0$,
the condition  $\varepsilon_- (M_M) = 0$
 suggests that a flattest energy band occurs at $a_0 = 0$.
Hence, by the previous result for $a_0$   at $\delta k_y = 0$,
degenerate perturbation theory at $M_M$ predicts a flat energy band when
\begin{equation}
v_D k_x = \sqrt{w_1^2 - w_0^{(0) 2}} .
\label{flt_bnd_cndtn}
\end{equation}
The magic twist angle is thereby obtained from the relation $k_x = K (\sin \alpha)$.
It yields the formula
\begin{equation}
2 \sin \alpha_* = \sqrt{w_1^2 - w_0^{(0) 2}} / (v_D K/2)
\label{flt_bnd_cndtn_bis}
\end{equation}
for the magic twist angle.

And what is the dispersion of the energy bands closest to zero energy along the bisecting $y$ axis?
Inspection of the coefficients (\ref{a_n})
yields that $a_1$ is linear order in $\delta k_y$,
while the remaining coefficients are quadratic in it.
The characteristic equation (\ref{chrctrstc_eq}) can be solved to first order in $a_1$ ($\delta k_y$):
$\varepsilon = \varepsilon^{(0)} + \varepsilon^{(1)}$.  
This yields the following expression for the first-order correction:
\begin{equation}
\varepsilon_{\pm}^{(1)} = - {a_1\over{2 (2 \varepsilon_{\pm}^{(0) 2} + a_2)}}.
\label{eps_1}
\end{equation}
Substituting in the result (\ref{eps_0}) for $\varepsilon_{\pm}^{(0) 2}$ yields
the result for the first-order correction:
\begin{equation}
\varepsilon_{\pm}^{(1)} = \mp {a_1\over{2\sqrt{a_2^2 - 4 a_0}}} .
\label{split}
\end{equation}
In turn, to lowest order in $\delta k_y$,
substituting in the coefficients (\ref{a_n}) $a_n$ of the characteristic equation above
yields the final result:
\begin{equation}
\varepsilon_{\pm}^{(1)} = \pm v_D (\delta k_y) .
\label{dsprsn_MM}
\end{equation}
The dispersion of the energy along the bisecting $y$ axis is then linear in $k_y$
 near the high-symmetry point $M_M$.

And as predicted by degenerate perturbation theory,
what is the nature of the energy spectrum near the opposing point in momentum,
$- M_M {\hat{\bi y}}$, which bisects the 
pair of Dirac nodes at $-{\bi d}_1(-)$ and $-{\bi d}_1(+)$ ?
These are depicted in Fig. \ref{dgnrt_prtrbtn_M_M} by the red arrows.
The truncated hamiltonian is once again a $4\times 4$ matrix of the form (\ref{H_MM+}),
but with $z_-$ and $z_+$ replaced by
\begin{eqnarray}
{\bar z}_- = - v_D e^{-i \alpha} i (k_{+,x} - i k_{+,y}) , \nonumber \\
{\bar z}_+ = - v_D e^{i \alpha} i (k_{-,x} - i k_{-,y}) ,
\label{z_-M_M}
\end{eqnarray}
and with $w_0$ replaced by $w_0^*$.
The matrix element $w_1$ is real, and it therefore remains unchanged.  
Inspection of the characteristic equation (\ref{chrctrstc_eq}) and (\ref{a_n}) then yields
that the energy eigenvalues are replaced by $\varepsilon \rightarrow -\varepsilon$.
This implies that the energy spectrum near zero energy and momentum $M_M$ shows particle-hole symmetry.
Furthermore, the first-order correction in $\delta k_y$ to the energy is minus that due to the opposite valley (\ref{dsprsn_MM}).
This implies that the dispersion of the lowest-energy bands {\it cross} at the high-symmetry point $M_M$
as a function of momentum along the bisecting $y$ axis.

We can now  answer the following question.
What is the energy dispersion as $\delta k_y$ gets farther away from the high-symmetry point $M_M$ ?
Consider $w_0^{(0)}$ as a perturbation instead of $\delta k_y$.
It is straight-forward to show that the energy eigenvalues (\ref{eps_0}) at $w_0^{(0)} = 0$ are
plus or minus
$|\varepsilon_{\pm}^{(0)}| = v_D\sqrt{k_x^2 +(\delta k_y)^2} \pm w_1$.
As expected, the degenerate energy from the Dirac cones is split by hopping in between the graphene sheets.
Further calculation yields the perturbation (\ref{eps_1})
$\varepsilon_{\pm}^{(1)} = \pm w_0^{(0)} [(\delta k_y) / \sqrt{k_x^2 + (\delta k_y)^2}]$.
To second order in $\delta k_y$, this yields approximate energy dispersions near the $M_M +$ point in momentum space:
\begin{eqnarray}
\varepsilon_{\pm,>} &\cong& + v_D k_x \pm w_1 - {1\over 2} {w_0^{(0) 2}\over{v_D k_x}}
+ {1\over 2} {1\over{v_D k_x}} [v_D (\delta k_y) \pm w_0^{(0)}]^2 , \nonumber \\
\varepsilon_{\pm,<} &\cong& - v_D k_x \mp w_1 + {1\over 2} {w_0^{(0) 2}\over{v_D k_x}}
- {1\over 2} {1\over{v_D k_x}} [v_D (\delta k_y) \mp w_0^{(0)}]^2 . \nonumber \\
\label{dsprsn_MM_more}
\end{eqnarray}
Likewise, the $M_M -$ point in momentum space yields minus the above dispersions:
\begin{eqnarray}
-\varepsilon_{\pm,>} &\cong& - v_D k_x \mp w_1 + {1\over 2} {w_0^{(0) 2}\over{v_D k_x}}
- {1\over 2} {1\over{v_D k_x}} [v_D (\delta k_y) \pm w_0^{(0)}]^2 , \nonumber \\
-\varepsilon_{\pm,<} &\cong& + v_D k_x \pm w_1 - {1\over 2} {w_0^{(0) 2}\over{v_D k_x}}
+ {1\over 2} {1\over{v_D k_x}} [v_D (\delta k_y) \mp w_0^{(0)}]^2 . \nonumber \\
\label{dsprsn_MM_minus}
\end{eqnarray}
Above, $\varepsilon_{\pm,>}$ in (\ref{dsprsn_MM_more}) and $-\varepsilon_{\pm,<}$ in (\ref{dsprsn_MM_minus})
are both positive, and they ``cross'' at $M_M$, while
$\varepsilon_{\pm,<}$ in (\ref{dsprsn_MM_more}) and $-\varepsilon_{\pm,>}$ in (\ref{dsprsn_MM_minus})
are both negative, and they also ``cross'' at $M_M$.
Notice that the ``cross'' dispersion at $M_M$ eventually turns away from zero energy as $\delta k_y$ gets large enough.
This is confirmed by Fig. \ref{dgnrt_prtrbtn_K&M} (b), where the roots of 
the characteristic equation, (\ref{chrctrstc_eq}) and (\ref{a_n}), are obtained numerically.


\section{Numerical Calculation of the Electron Energy Spectrum}
Below, we find the energy eigenvalues of the nearest-neighbor electron hopping model for
twisted bi-layer graphene that was introduced in subsection \ref{e_hppng_hmltnn}.
Commensurate twist angles near $1.1^{\circ}$ shall be studied.
(See Table \ref{near_magic}.)
Four flat bands near zero energy emerge at ``magic'' values of the matrix elements 
for hopping in between the twisted sheets of graphene.

\subsection{Hamiltonian Matrix}
The nearest neighbor electron hopping hamiltonian for bi-layer graphene at commensurate twist angles
was constructed in section \ref{e_hppng_hmltnn} in momentum space.
It has the form (\ref{mtrx_frm}),
with intra-graphene matrix elements $H_{--}$ and $H_{++}$
 given by (\ref{h_parallel}) and (\ref{zeta}),
and with inter-graphene matrix elements $H_{-+}$ and $H_{+-}$
 given by (\ref{h-+}) and the hermitian conjugate.
The $3$-fold rotation symmetry of twisted bi-layer graphene shown by Fig. \ref{cmmnsrt_tblg_7}, for example,
can be exploited to compute the latter.
Such calculations are described in \ref{ppndx_c_hmltn}.
Last, a gauge transformation that moves the phase factors from the inter-graphene-sheet A-B
matrix elements (\ref{h-+}) to the A-A(B-B) ones is made.  It is described in \ref{ppndx_a}.

\begin{table}
\begin{center}
\begin{tabular}{|c|c|c|c|c|}
\hline
$n$ & $m$ & $l$ & $\theta$ & $v_D K_M / 2 = (\pi/\sqrt{3 l}) t_{\parallel}$ \\
\hline
$29$ & $30$ & $2611$ & $1.121^{\circ}$ & $0.0355\, t_{\parallel}$ \\
$30$ & $31$ & $2791$ & $1.085^{\circ}$ & $0.0343\, t_{\parallel}$ \\
$31$ & $32$ & $2977$ & $1.050^{\circ}$ & $0.0332\, t_{\parallel}$ \\
\hline
\end{tabular}
\end{center}
\caption{Integer parameters for bi-layer graphene at commensurate twist angles $\theta$
along the Periodic Moir\'e Pattern line that are near the experimental magic twist angle.
The last column is the prediction by zone-folding
of the energy at the $M_M$ point of the first positive-energy band.}
\label{near_magic}
\end{table}

The dimensions of the hopping hamiltonian matrix are $4 l \times 4 l$,
where $l$ is the number of A (B) sites inside of a true moir\' e unit cell per sheet of graphene (\ref{l_M}).
Energy eigenvalues of the hamiltonian matrix are obtained by using standard numerical routines
for dense matrices.  We shall restrict the numerical calculations to commensurate lattices
along the Periodic Moir\'e Pattern line, $m = n+1$, that are closest to the experimental
magic twist angle of $\theta = 1.1^{\circ}$.
These are listed in Table \ref{near_magic}.
They result in the smallest possible dimensions for the hopping hamiltonian matrix.

\begin{figure}
\includegraphics[scale=1.00, angle=0]{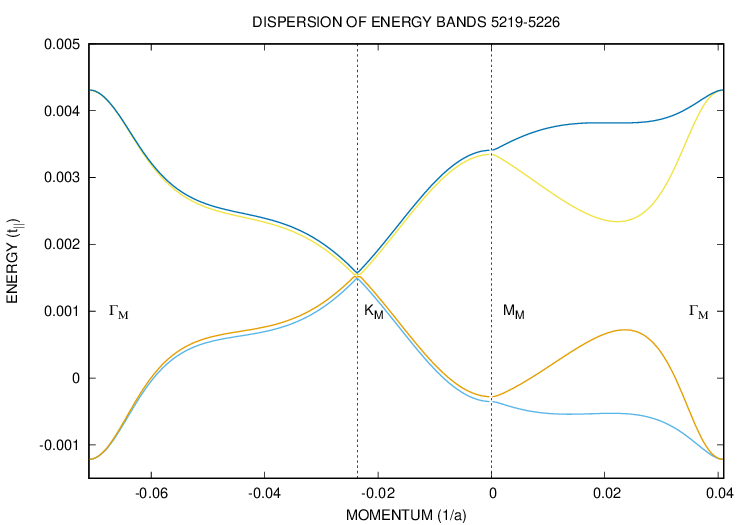}
\caption{Dispersion of flat energy bands near zero energy 
for hopping hamiltonian in momentum space (\ref{mtrx_frm}) of dimensions $4 l\times 4 l$,
 with $l = 2611$ A(B) sites per sheet of graphene inside a moir\'e unit cell.
(See Table \ref{near_magic}.)  Inter-graphene-sheet matrix elements
are uniform over nearest neighbors, and they are set to
$t_{\perp}^{(1)} = 0.069\, t_{\parallel} = t_{\perp}^{(0)}$.}
\label{ideal_inter_sheet_hops}
\end{figure}

\subsection{Ideal Nearest Neighbor Hopping Between Twisted Sheets of Graphene}
Figure \ref{ideal_inter_sheet_hops} shows the flat bands of the energy spectrum near the magic twist angle
for integer parameters $n=29$, $m=30$, and $l = 2611$.
The hopping matrix elements in between graphene sheets are set to
$t_{\perp}^{(1)} = 0.069\, t_{\parallel} = t_{\perp}^{(0)}$
across all nearest neighbor links.  These are depicted, respectively,
by Figs. \ref{twist_A_B} and \ref{twist_A_A}. 
By comparison, the bandwidth of these flat bands reaches a minimum at inter-graphene-sheet
matrix elements $t_{\perp}^{(1)} = 0.072\, t_{\parallel} = t_{\perp}^{(0)}$.
Notice that the difference in energy between the $M_M$ point and the $K_M$ point
in Fig. \ref{ideal_inter_sheet_hops}
is an order of magnitude smaller than that predicted by zone folding in Table \ref{near_magic}.
The level crossing along $\Gamma_M$-$M_M$-$\Gamma_M$ at the $M_M$ point that 
is predicted by degenerate perturbation theory, Fig. \ref{dgnrt_prtrbtn_K&M} b, is evident.
There is a small amount of level repulsion there too, however.
It is a result of inter-valley mixing.
And as predicted by degenerate perturbation theory, Fig. \ref{dgnrt_prtrbtn_K&M} a,
the results of inter-valley mixing are also visible at the $K_M$ point.
Last, the remaining two cases of commensurate magic twist angles listed in Table \ref{near_magic} give similar
results for the flat energy bands at $t_{\perp}^{(0)} = t_{\perp}^{(1)}$.

\begin{figure}
\includegraphics[scale=1.00, angle=0]{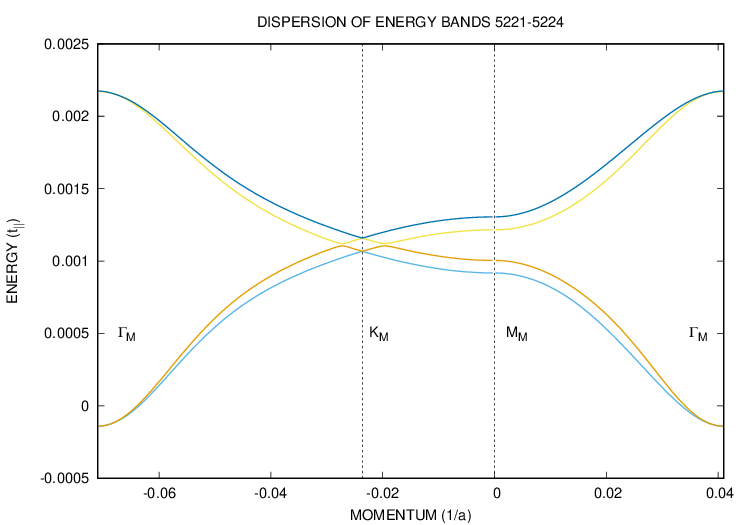}
\caption{Flat energy bands near zero energy for hopping hamiltonian (\ref{mtrx_frm}),
with $l = 2611$ A(B) sites per sheet of graphene inside a moir\'e unit cell.
(See Table \ref{near_magic}.)  ``Relaxed'' hopping matrix elements
are uniform over nearest neighbors between the
graphene sheets, and they are set to
$t_{\perp}^{(1)} = 0.079\, t_{\parallel}$ and to $t_{\perp}^{(0)} = 0.0395\, t_{\parallel}$.}
\label{relax_inter_sheet_hops}
\end{figure}

\subsection{``Relaxed'' Nearest Neighbor Hopping Between Twisted Sheets of Graphene}
The bi-layer graphene crystals at commensurate twist angles considered in this paper
have honeycomb moir\'e unit cells with Bernal stacks at the vertices. 
Figure \ref{cmmnsrt_tblg_7} displays this in the case of the smallest commensurate
twisted bi-layer graphene crystal, with integer parameters 
$m=2$, $n=1$, and $l = 7$.
In addition, the center of such honeycomb moir\'e unit cells coincides 
with a twist center for A-A links in between the two sheets of graphene.
Given the small twist angle of $1.1^{\circ}$ that is experimentally observed
in twisted bi-layer graphene,
A(B) sites in between the two sheets of graphene
will lie practically right above/below
each other in the vicinity of such twist centers.
It has been pointed out that strain will
develop in this region of the moir\'e unit cell
in the ideal twisted bi-layer graphene crystal
so that nearest neighbor links of A(B)
sites in between the two sheets of graphene 
become more separated in the intra-graphene
directions\cite{carr_prb_18}\cite{yoo_rcnstrct_tblg_2019}\cite{guinea_walet_19}\cite{carr_kaxiras_prr_19}\cite{walet_guinea_20}\cite{leconte_prb_22}.
Below, we shall model this effect by assuming
uniform hopping between nearest neighbor A(B) sites in between graphene sheets, $t_{\perp}^{(0)}$,
and  by assuming
uniform hopping between nearest neighbor A sites in one sheet of graphene 
with a B site in the other sheet of graphene, $t_{\perp}^{(1)}$,
but while satisfying the inequality $t_{\perp}^{(0)} < t_{\perp}^{(1)}$.
By the previous, the ratio $t_{\perp}^{(0)} / t_{\perp}^{(1)}$ coincides with the
ratio of matrix elements at the $M_M$ point in momentum space,
$w_{A-,A+} / w_{A-,B+}$.
It coincides with the reduction factor for the A-A(B-B) links versus the A-B links
in the moire patterns of twisted bi-layer graphene\cite{vHs_tblg_nature_19b}.

Figure \ref{relax_inter_sheet_hops} shows the ``relaxed'' flat bands of
 the energy spectrum near the magic twist angle
for integer parameters $n=29$, $m=30$, and $l = 2611$.
The hopping matrix elements in between graphene sheets are set to
$t_{\perp}^{(1)} = 0.079\, t_{\parallel}$ and to $t_{\perp}^{(0)} = 0.0395\, t_{\parallel}$,
which is half of the former,
across all nearest neighbor links. 
This choice of inter-graphene-sheet hopping matrix elements 
therefore lies halfway between the previous ideal case
and the chiral limit, $t_{\perp}^{(0)} = 0$,
in which case the central bands are perfectly flat\cite{chrl_lmt_tblg_prl_19}.
Figures \ref{twist_A_B} and \ref{twist_A_A}
 depict these inter-graphene-sheet links, respectively.
By comparison, the bandwidth of these ``relaxed'' flat bands reaches a minimum at inter-graphene-sheet
matrix elements $t_{\perp}^{(1)} = 0.078\, t_{\parallel}$ and  $t_{\perp}^{(0)} = 0.039\, t_{\parallel}$,
which again is half the former.
Figure \ref{relax_inter_sheet_hops} shows that the splitting of valley degeneracy
exists in general at all momenta.  Also, the difference between the energy at the $M_M$ point 
and the energy at the $K_M$ is two orders of magnitude smaller than that predicted by zone-folding
in Table \ref{near_magic}.
Again, the remaining two cases of commensurate magic twist angles listed in Table \ref{near_magic} give similar
results for the flat energy bands  at $t_{\perp}^{(0)} = t_{\perp}^{(1)} / 2$.

\subsection{Magic Twist Angle versus Degenerate Perturbation Theory}
By formula (\ref{flt_bnd_cndtn}),
degenerate perturbation theory predicts that flat electron bands emerge
when the energy at $M_M$ predicted by zone folding in the first positive-energy band
is equal to $\sqrt{w_1^2 - w_0^2}$.
In the limit of large integer parameters along the Periodic Moir\'e Pattern line, $m = n + 1$,
degenerate perturbation theory also predicts that 
$w_1 = - 0.5\, t_{\perp}^{(1)}$ and $w_0 = - 0.5\, t_{\perp}^{(0)}$.
Recall that the ``relaxed'' flat bands depicted by Fig. \ref{relax_inter_sheet_hops}
occur near an optimally flat point at 
$t_{\perp}^{(1)} = 0.078\, t_{\parallel}$ and  $t_{\perp}^{(0)} = t_{\perp}^{(1)} / 2$.
Hence, by the previous,
$\sqrt{w_1^2 - w_0^2} = (\sqrt{3}/2) w_1 = 0.0338 \, t_{\parallel}$.
This calculation compares well with the result from zone-folding,
$v_D K_M / 2 = 0.0355 \, t_{\parallel}$,
that is listed in Table \ref{near_magic}
at the corresponding twist angle,
which is what is predicted by degenerate perturbation theory (\ref{flt_bnd_cndtn}).

Figure \ref{ideal_inter_sheet_hops}, on the other hand,
shows flat bands at ideal inter-graphene-sheet hopping.
It lies near an optimally flat point at
$t_{\perp}^{(1)} = 0.072\, t_{\parallel} = t_{\perp}^{(0)}$.
In the ideal case, however,
 we generally have that
$\sqrt{w_1^2 - w_0^2} \cong 0$
 at large integer parameter $n$.
It thereby compares badly with the answer from zone-folding that is listed in Table \ref{near_magic},
which is what is predicted by degenerate perturbation theory (\ref{flt_bnd_cndtn}).

\section{Discussion}
Below, we compare the  bi-layer graphene lattice at commensurate twist angles studied here
to the Bistritzer-MacDonald twist construction of the reciprocal lattice\cite{bistritzer_macdonald_11}.
We also reveal the connection between inter-valley mixing (\ref{w_1_prime}) and spiral Kekul\'e order.

\subsection{Moir\'e Patterns with Subcells in Twisted Bi-Layer Graphene}
The bi-layer graphene lattices studied above in section \ref{cmmnsrt_tblg} 
have commensurate twist angles 
characterized by relatively prime integer parameters\cite{campanera_prb_07}\cite{trambly_nano_lett_10} $m > n$,
where half the twist angle $\alpha$ is given by (\ref{sine}) and (\ref{l_M}).
It is instructive to compute the difference between Dirac nodes in the minus-rotated sheet of graphene
and the plus-rotated sheet of graphene:
\begin{equation}
{\bi d}_1 (-) - {\bi d}_1 (+) = (m-n) {\bi D}_1
\label{twst_drc_nds}
\end{equation}
by (\ref{d1}).
At $m = n+1$, the twist in the Dirac nodes goes between nearest-neighbor moir\'e Dirac nodes:
i.e., in between the dashed lines in Fig. \ref{rcprcl_lttc_a}.
The series of lattices of twisted bi-layer graphene that are constrained by $m-n = 1$
is therefore consistent with the twist construction for the moir\'e reciprocal lattice
introduced by Bistritzer and MacDonald \cite{bistritzer_macdonald_11}.
In that construction,
the displacement (\ref{twst_drc_nds}) coincides with a segment of a honeycomb cell
that is at the edge of the first moir\'e Brillouin zone.
Recall that in this case,
the cells of the moire pattern in real space are strictly periodic.

\begin{figure}
\includegraphics[scale=1.00, angle=0]{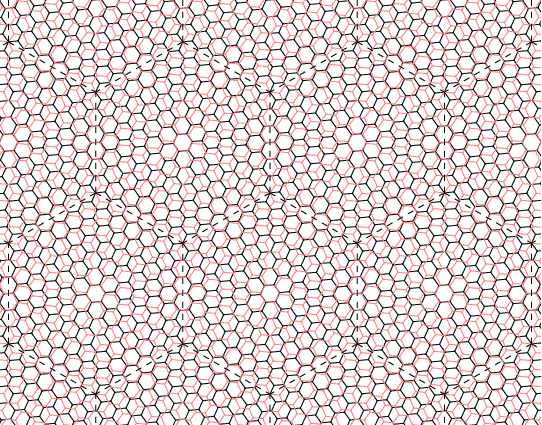}
\caption{Moir\'e pattern of two honeycombs twisted 
by a commensurate angle $\theta = 10.99^{\circ}$.
It corresponds to integer parameters $m=7$ and $n=5$.
The dashed lines mark the edges of the {\it true} Wigner-Seitz unit cell,
with $l = 109$ A(B) sites per sheet of graphene inside of it.}
\label{moire_pattern_109}
\end{figure}

At first sight,
on the other hand,
bi-layer graphene with commensurate twist angles at integer parameters $m-n > 1$
is  not consistent with 
the Bistritzer-MacDonald twist construction of the reciprocal lattice\cite{bistritzer_macdonald_11}.
By Fig. \ref{rcprcl_lttc_a} and (\ref{twst_drc_nds}),
 applying equal and opposite twists of a Dirac node in graphene by an angle $\alpha$
about the $\Gamma$ point
 results in honeycomb supercells in momentum space
that include $(m-n)^2$ moir\'e reciprocal lattice vector in this case.
The  Bistritzer-MacDonald twist construction in this general case remains physically relevant, however.
Applying duality to the $(m-n)^2$ moir\'e Brillouin zones in the honeycomb supercell 
indicates that the true Wigner-Seitz unit cell of the bi-layer graphene lattice
contains $(m-n)^2$ subcells corresponding to each cell of the moir\'e pattern.
Figure \ref{moire_pattern_109} displays the moir\'e pattern for integer parameters $m=7$ and $n=5$.
It reveals a moir\'e subcell at the center of the true Wigner-Seitz unit cell,
plus moir\'e subcells at the center of the edges of the true Wigner-Seitz unit cell.
This results, correctly, in $1 + {1\over 2} 6 = 4$ moir\'e subcells in each true Wigner-Seitz unit cell.
The moir\'e subcells are notably {\it not} periodic.

\subsection{Inter-Valley Mixing versus Kekul\'e Spiral Order}
The truncated hamiltonian (\ref{H_KM+}) for an electron carrying momentum near the corner of the moir\'e Brillouin zone
 reveals that the $K+$ and $K-$ valleys in the different sheets of graphene
mix according to the matrix elements
$w^{\prime}_{A-,B+} = w_1^{\prime}$ and $w^{\prime}_{A-,A+} = w_0^{\prime}$.
Along the Periodic Moire Pattern line, $m = n+1$, nearest-neighbor hopping between the two graphene sheets
is described by Figs. \ref{twist_A_B} and \ref{twist_A_A}.  
Following  expressions (\ref{h_perp_AB})-(\ref{dot_products_K_M_a}),
this  yields the formula (\ref{w_1_prime}) for the matrix element $w^{\prime}_{A-,B+}$.
It is more generally given by
\begin{equation}
w_1^{\prime} = -
\sum_{(n_1,n_2)\in {\rm mWSuc}} {t_{\perp}^{(1)}\over l} (e^{i 2\pi/3})^{n_1+n_2} ,
\label{w_1_prime_bis}
\end{equation}
where $n_1 {\bi a}_1 + n_2 {\bi a}_2$ lie inside the moir\'e Wigner-Seitz unit cell (mWSuc).
The matrix element $w_1^{\prime}$
is thereby small and of order ${t_{\perp}^{(1)}/ l}$ due to the summation of phase factors
that  are cube roots of unity.

The field of cube roots of unity $(e^{i 2\pi/3})^{n_1+n_2}$ in formula (\ref{w_1_prime_bis})
for the inter-valley matrix element $w^{\prime}_{A-,B+}$ describes a Kekul\'e spiral
over the honeycomb cells of a sheet of graphene.  Recent theoretical work predicts incommensurate
Kekul\'e spiral order in magic-angle twisted bi-layer graphene at a filling of
$\nu = 2$ electrons or holes per moir\'e unit cell\cite{kwan_prx_21}.
Recent scanning-tunnelling microscopy (STM) studies of magic-angle twisted bi-layer graphene
have confirmed this prediction\cite{ivc_tblg_nature_23}.
Further, recent STM studies of trilayer graphene at the magic twist angle
also  find evidence for Kekul\'e spirals\cite{ivc_ttlg_nature_23}.
Such naturally ocurring Kekul\'e spiral order could therefore result in a
dramatic increase in the inter-valley mixing according to formula (\ref{w_1_prime_bis})
for the matrix element $w^{\prime}_{A-,B+}$.

\section{Conclusions}
We have studied a nearest-neighbor electron hopping model for twisted bi-layer graphene.
A commensurate twist about a Bernal stack is made, resulting in a super-honeycomb of Bernal stacks.
(See Fig. \ref{cmmnsrt_tblg_7}.)
A mechanism for flat central bands at the magic twist angle
is revealed.  It is due to optimal level repulsion 
at the middle of the edge of the moir\'e Brillouin zone.
This result was obtained in the case of periodic moir\'e patterns with no subcells,
in the limit that they are large.

The nearest-neighbor electron hopping model also shows appreciable breaking of valley degeneracy
of the flat bands near the magic twist angle.
In particular, Fig. \ref{relax_inter_sheet_hops} shows
substantial breaking of valley degeneracy at the $M_M$ point in momentum space,
which corresponds to the van Hove singularity.
It could lead to correlated insulating states at electron filling such that the chemical potential
lies within the split energies there.  This will be investigated in future studies.

\subsection{Acknowledgments}
The author thanks Eslam Khalaf and Christopher Gutierrez for useful conversations.

\clearpage

\appendix\section{Gauge Transformation\label{ppndx_a}}
The phase factors associated with the off-diagonal A/B and B/A matrix elements (\ref{h-+})
can be removed by a gauge transformation.  
In particular, make the following replacement of the states in the 
entire hopping hamiltonian,
including the diagonal parts, $H_{--}$ and $H_{++}$,
and the off-diagonal parts, $H_{-+}$ and $H_{+-}$:
\begin{eqnarray}
|{\bi k}-{\bi G}_i,B,-\rangle\rangle &\rightarrow & 
e^{+i{\bi G}_i\cdot{\bi C}_2} |{\bi k}-{\bi G}_i,B,-\rangle\rangle , \nonumber\\
|{\bi k}-{\bi G}_j,A,+\rangle\rangle &\rightarrow & 
e^{+i{\bi G}_j\cdot{\bi C}_2} |{\bi k}-{\bi G}_j,A,+\rangle\rangle ,
\label{A_B_g_t}
\end{eqnarray}
and 
\begin{eqnarray}
|{\bi k}-{\bi G}_i,A,-\rangle\rangle &\rightarrow & 
e^{-i{\bi G}_i\cdot{\bi C}_3} |{\bi k}-{\bi G}_i,A,-\rangle\rangle , \nonumber\\
|{\bi k}-{\bi G}_j,B,+\rangle\rangle &\rightarrow & 
e^{-i{\bi G}_j\cdot{\bi C}_3} |{\bi k}-{\bi G}_j,B,+\rangle\rangle .
\label{B_A_g_t}
\end{eqnarray}
This results in the replacement
\begin{equation}
\zeta_{\pm}({\bi k} - {\bi G}_i)\rightarrow e^{\mp i{\bi G}_i\cdot{\bi C}_1} \zeta_{\pm}({\bi k} - {\bi G}_i)
\label{rplc_zt}
\end{equation}
of the phase factors for the intra-graphene-sheet matrix elements (\ref{h_parallel}),
and in the replacements
\begin{eqnarray}
\langle\langle {\bi k}_i,A,-|H|{\bi k}_j,A,+\rangle\rangle &=& 
-e^{-i({\bi G}_i\cdot{\bi C}_3+{\bi G}_j\cdot{\bi C}_2)} \cdot \nonumber \\
&& \cdot \sum_{\delta{\bi r}_{A/A}}{t_{\perp}^{(0)}\over l}
e^{i[-{\bi k}_i\cdot\delta{\bi r}_{A/A}(-) + {\bi k}_j\cdot\delta{\bi r}_{A/A}(+)]} \nonumber\\
%
%
%
\langle\langle {\bi k}_i,A,-|H|{\bi k}_j,B,+\rangle\rangle &=& -
\sum_{\delta{\bi r}_{B/A}} {t_{\perp}^{(1)}\over l}
e^{i[-{\bi k}_i\cdot\delta{\bi r}_{B/A}(-) + {\bi k}_j\cdot\delta{\bi r}_{B/A}(+)]} \nonumber\\
%
%
%
\langle\langle {\bi k}_i,B,-|H|{\bi k}_j,A,+\rangle\rangle &=& -
\sum_{\delta{\bi r}_{A/B}} {t_{\perp}^{(1)}\over l}
e^{i[-{\bi k}_i\cdot\delta{\bi r}_{A/B}(-) + {\bi k}_j\cdot\delta{\bi r}_{A/B}(+)]} \nonumber\\
%
%
%
\langle\langle {\bi k}_i,B,-|H|{\bi k}_j,B,+\rangle\rangle &=& 
-e^{+i({\bi G}_i\cdot{\bi C}_2+{\bi G}_j\cdot{\bi C}_3)} \cdot \nonumber \\
&& \cdot \sum_{\delta{\bi r}_{B/B}}{t_{\perp}^{(0)}\over l}
e^{i[-{\bi k}_i\cdot\delta{\bi r}_{B/B}(-) + {\bi k}_j\cdot\delta{\bi r}_{B/B}(+)]} \nonumber\\
\label{hmp}
\end{eqnarray}
of the inter-graphene-sheet matrix elements (\ref{h-+}).
In the right-hand side of (\ref{rplc_zt}),
we have used the identity ${\bi C}_1 + {\bi C}_2 + {\bi C}_3 = 0$.
Above, also,
${\bi k}_i = {\bi k} - {\bi G}_i$ and ${\bi k}_j = {\bi k} - {\bi G}_j$.
Notice that this guage transformation consolidates the phase factors
on the block-diagonal intra-graphene parts of the hopping hamiltonian
in the chiral limit\cite{chrl_lmt_tblg_prl_19},
where there is no $A/A$ nor $B/B$ inter-graphene-sheet hopping.

\section{Phase Factors, Quotient Groups\label{ppndx_b}}
The product of the phase factor $e^{i{\bi k}\cdot{\bi c}_1(\pm)}$ with (\ref{zeta_prime})
$\zeta_{\pm}^{\prime}({\bi k})$ gives the intra-graphene matrix element (\ref{zeta}) $\zeta_{\pm}({\bi k})$.
The reduced matrix element $\zeta_{\pm}^{\prime}({\bi k})$ is periodic over the graphene Brillouin zone,
while the phase factor $e^{i{\bi k}\cdot{\bi c}_1 (\pm)}$ is not.
Let us compute the  phase factor in twisted bi-layer graphene at momentum ${\bi k} = + K_{\widebar{M}} {\hat{\bi x}}$,
where the $K+$ Dirac zero of the $+\alpha$-rotated sheet of graphene coincides with
the $K-$ Dirac zero of the $-\alpha$-rotated sheet of graphene. 
(See Fig. \ref{g_twst_7}, red dot in black circle.)
In the latter case,
the argument of the exponential is imaginary $i$ times the dot product
$K_{\widebar{M}} {\hat{\bi x}} \cdot {\bi c}_1(-) = K_{\widebar{M}} {\hat{\bi x}} (+) \cdot {\bi c}_1$,
where the notation ${\bi v} (\pm)$ means that the vector ${\bi v}$ is rotated by $\pm \alpha$.
Now observe that $K_{\widebar{M}} {\hat{\bi x}} (+) = -m {\bi d}_2 + n {\bi d}_3$.
By (\ref{d_i}), we thereby obtain
$(K a) (m+n){\sqrt{3}\over 2} = 2\pi (m+n)/3$ times imaginary $i$ for the argument of the exponential, or
\begin{equation}
e^{+iK_{\widebar{M}} {\hat{\bi x}}\cdot{\bi c}_1(-)} = e^{i 2\pi (m+n)/3}.
\label{phase_factor_c-}
\end{equation}
In the case of the phase factor $e^{+iK_{\widebar{M}} {\hat{\bi x}}\cdot{\bi c}_1(+)}$, on the other hand,
the argument of the exponential is imaginary $i$ times the dot product
$K_{\widebar{M}} {\hat{\bi x}} \cdot {\bi c}_1(+) = K_{\widebar{M}} {\hat{\bi x}} (-) \cdot {\bi c}_1$.
Observe, once more, that
$K_{\widebar{M}} {\hat{\bi x}} (-) = -n {\bi d}_2 + m {\bi d}_3$.
Compared to the previous case, we have swapped $m$ with $n$.  
The phase factor in this case
is thereby given by the phase factor in the previous case (\ref{phase_factor_c-}).

The other phase factor that appears in the nearest-neighbor electron hopping model
for bi-layer graphene at commensurate twist angles is linked to the inter-graphene-sheet
matrix elements between A sites and B sites (\ref{h-+}).
In the case of hopping from a
 B site in the plus-rotated sheet of graphene to an A site in the minus-rotated sheet of graphene,
for example, the phase factor is given by
$e^{i {\bi G}_-\cdot {\bi C}_3} e^{-i {\bi G}_+\cdot {\bi C}_3}$,
where ${\bi G}_- = n_1^{(-)} {\bi B}_1 + n_2^{(-)} {\bi B}_2$ 
and ${\bi G}_- = n_1^{(+)} {\bi B}_1 + n_2^{(+)} {\bi B}_2$
are  moir\'e reciprocal lattice vectors.
Using expressions (\ref{C_i}) and (\ref{B_i}) for the moir\'e A-B displacements ${\bi C}_i$ 
(see Fig. \ref{cmmnsrt_tblg_7})
and the primitive vectors ${\bi B}_j$ of the moir\'e reciprocal lattice,
we get the result
$e^{i {\bi G}_{\pm}\cdot {\bi C}_1} = (e^{-i 2\pi/3})^{n_1^{(\pm)} + n_2^{(\pm)}}$.
Next, using the identity ${\bi C}_3 = {\bi C}_1 + {\bi C}_3 - {\bi C}_1 = {\bi C}_1 - {\bi A}_2$,
we get the same result for $e^{i {\bi G}_{\pm}\cdot {\bi C}_3}$.
The net phase factor is therefore
\begin{equation}
e^{i {\bi G}_-\cdot {\bi C}_3} e^{-i {\bi G}_+\cdot {\bi C}_3} =
(e^{-i 2\pi/3})^{n_1^{(-)} + n_2^{(-)}} (e^{+i 2\pi/3})^{n_1^{(+)} + n_2^{(+)}} .
\label{other_phase_factor}
\end{equation}
Importantly, this result yields that the net phase factor is {\it unity} if
$(n_1^{(-)} + n_2^{(-)})$ mod $3 = (n_1^{(+)} + n_2^{(+)})$ mod $3$.
In other words, the net phase factor (\ref{other_phase_factor}) is unity when
${\bi G}_-$ and ${\bi G}_+$ lie in the {\it same}
 $3$-fold sublattice of the triangular moir\'e reciprocal lattice.
These are (a) sites with $(n_1 + n_2)$ mod $3 = 1$,
(b) sites with $(n_1 + n_2)$ mod $3 = 2$, and
(c) sites with $(n_1 + n_2)$ mod $3 = 0$.
Such a and b sites form a honeycomb on the original triangular moir\'e reciprocal lattice,
with c sites at the centers of the honeycomb cells.

Last, Fig. \ref{rcprcl_lttc_b} shows the first Brillouin zones of
 the plus-rotated and the minus-rotated sheets of graphene,
$1BZ(+)$ and $1BZ(-)$, respectively.
Notice that both Brillouin zones contain 
the {\it same} moir\'e reciprocal lattice vectors, $\{ {\bi G}_0, {\bi G}_1, ..., {\bi G}_{l-1} \}$, 
with $l = 61$ in this particular case.
The periodic structure of the first Brillouin zone $1BZ(+)$ displayed by Fig. \ref{rcprcl_lttc_b} 
means that a {\it quotient group} exists that is  composed of the $l$ moir\'e
reciprocal lattice vectors that it contains.
Summation of two elements in the group is defined up to the periodicity:
${\bi G}_+ + {\bi G}_+^{\prime} \rightarrow ({\bi G}_+ + {\bi G}_+^{\prime})$ mod $1BZ(+)$.
The quotient group is therefore equal to 
$Z^2 ({\rm moir\acute{e}}) / Z^2 ({\rm graphene} +)$,
where $Z^2 ({\rm moir\acute{e}})$ is the moir\'e reciprocal lattice,
and where $Z^2 ({\rm graphene} +)$ is the reciprocal lattice of the plus-rotated sheet of graphene.
A related group exists based on the minus-rotated sheet of graphene,
where $1BZ(+)$ is replaced by $1BZ(-)$. (See Fig. \ref{rcprcl_lttc_b}.)

The above quotient groups emerge when momentum is added to  bi-layer graphene at commensurate twist angles.
In particular, notice that the difference between
$({\bi G}_+ + \Delta {\bi G})$ mod $1BZ(+)$ and
$({\bi G}_- + \Delta {\bi G})$ mod $1BZ(-)$
is {\it not} necessarily equal to the difference between ${\bi G}_+$ and ${\bi G}_-$.
This is a result of the twist angle.
However, the previous phase factor (\ref{other_phase_factor}) does indeed remain the same:
\begin{equation}
e^{i ({\bi G}_- + \Delta {\bi G})\cdot {\bi C}_3} e^{-i ({\bi G}_+ + \Delta {\bi G})\cdot {\bi C}_3} =
e^{i {\bi G}_-\cdot {\bi C}_3} e^{-i {\bi G}_+\cdot {\bi C}_3} .
\label{same_phase_factor}
\end{equation}
This property can be exploited when computing the previous phase factor after the addition of a moir\'e
reciprocal lattice vector $\Delta {\bi G}$.
(See Fig. \ref{dgnrt_prtrbtn_M_M} and the end of subsection \ref{dgnrt_prtrbtn_thry_M_M}.)

\section{Inter-Graphene Matrix Elements: $3$-Fold Rotation Symmetry, Geometric Series Sums\label{ppndx_c}}
Below,
we shall compute explicitely the inter-graphene-sheet matrix elements that appear in
the four-fold degenerate perturbation theory analysis found in section \ref{dgnrt_prtrbtn}.
In particular,
we shall confine ourselves to twisted bi-layer graphene
at integer parameters $m = n+1$.
The nearest-neighbor hopping links in between the two sheets of graphene
have the ``hexagon'' form depicted by Figs. \ref{twist_A_B} and \ref{twist_A_A} in such case.
Also,
the nearest-neighbor hopping model hamiltonian
has the form (\ref{mtrx_frm}),
with matrix elements (\ref{h_parallel}) and (\ref{h-+}) in momentum space
for $H_{--}$, $H_{-+}$, etc. .
The simplest hopping model shall be assumed:
{\it all} of the nearest-neighbor hopping matrix elements
between the twisted graphene sheets are unique, and they are given by $-t_{\perp}^{(0)}$ and $-t_{\perp}^{(1)}$.
Here, the superscript $0$ refers to A-A links between sheets of graphene,
while the superscript $1$ refers to the corresponding A-B links.

\subsection{Application of $3$-Fold and $6$-Fold Rotation Symmetry}\label{ppndx_c_hmltn}
Figure \ref{cmmnsrt_tblg_7} reveals that the twist centers at ${B+}/{A-}$ Bernal stacks
show $6$-fold rotation symmetry.  It becomes useful to exploit this symmetry when computing
the corresponding matrix element (\ref{h-+})
$\langle\langle {\bi k}_i,A,-|H|{\bi k}_j,B,+\rangle\rangle$.
In particular, the sum over inter-graphene links can be re-expressed as
\begin{eqnarray}
h_{A-,B+} = {t_{\perp}^{(1)}\over l} \biggl[1 + 
2\, {\rm Re} \sum_{n_1=1}^{n}\sum_{n_2=0}^{n_1 - 1} 
(& e^{i[-{\bi k}_i\cdot\delta{\bi r}_{B/A}(-) + {\bi k}_j\cdot\delta{\bi r}_{B/A}(+)]} + \nonumber \\
 & e^{i[-{\bi k}_i\cdot\delta{\bi r}_{B/A}^{\prime}(-) + {\bi k}_j\cdot\delta{\bi r}_{B/A}^{\prime}(+)]} + \nonumber \\
 & e^{i[-{\bi k}_i\cdot\delta{\bi r}_{B/A}^{\prime\prime}(-) + {\bi k}_j\cdot\delta{\bi r}_{B/A}^{\prime\prime}(+)]})\biggr] ,
\label{h_A-_B+}
\end{eqnarray}
with $3$-fold-symmetric displacements from the Bernal stack
\begin{eqnarray}
\delta{\bi r}_{B/A}(\pm) = n_1 {\bi a}_1(\pm) + n_2 {\bi a}_2(\pm) , \nonumber \\
\delta{\bi r}_{B/A}^{\prime}(\pm) = n_1 {\bi a}_2(\pm) + n_2 {\bi a}_3(\pm) , \nonumber \\
\delta{\bi r}_{B/A}^{\prime\prime}(\pm) = n_1 {\bi a}_3(\pm) + n_2 {\bi a}_1(\pm) .
\label{dr_B/A's}
\end{eqnarray}
The phase factors and minus signs in (\ref{h-+}) have been suppressed above in (\ref{h_A-_B+}).
Figure \ref{twist_A_B} shows the ${B+}/{A-}$ links between
 the twisted sheets of graphene within a moir\'e Wigner-Seitz cell.
The sums over the integers $n_1$ and $n_2$ in (\ref{h_A-_B+})
represent the sector including $90^\circ$ up from the twist center to $150^\circ$ from the twist center.

Figure \ref{cmmnsrt_tblg_7} also reveals that the twist centers of ${A+}/{A-}$ links between twisted sheets of graphene
show instead only $3$-fold rotation symmetry.
This symmetry can once again be exploited, however, to compute  
the corresponding matrix element (\ref{h-+})
$\langle\langle {\bi k}_i,A,-|H|{\bi k}_j,A,+\rangle\rangle$.
Figure \ref{twist_A_A} displays the ${A+}/{A-}$ links between the twisted sheets of graphene
inside a moir\'e Wigner-Seitz cell.
Notice the nucleus of $6$ links in the inner triangle, 
plus $3$ trapezoids adjacent to the inner triangle, 
in addition to $3$ outer triangles adjacent to the trapezoids.
Also observe that a trapezoid plus an adjacent outer triangle form a parallelogram!
By exploiting $3$-fold rotation symmetry,
the sum on the right-hand side of the expression (\ref{h-+}) for the matrix element
$\langle\langle {\bi k}_i,A,-|H|{\bi k}_j,A,+\rangle\rangle$
becomes
\begin{eqnarray}
h_{A-,A+}^{({\rm ncls})} = {t_{\perp}^{(0)}\over l} \sum_{(m_1,m_2)=(0,-2),(1,0)} 
(& e^{i[-{\bi k}_i\cdot\delta{\bi r}_{A/A}(-) + {\bi k}_j\cdot\delta{\bi r}_{A/A}(+)]} + \nonumber \\
 & e^{i[-{\bi k}_i\cdot\delta{\bi r}_{A/A}^{\prime}(-) + {\bi k}_j\cdot\delta{\bi r}_{A/A}^{\prime}(+)]} + \nonumber \\
 & e^{i[-{\bi k}_i\cdot\delta{\bi r}_{A/A}^{\prime\prime}(-) + {\bi k}_j\cdot\delta{\bi r}_{A/A}^{\prime\prime}(+)]}) ,
\label{h_A-_A+_ncls}
\end{eqnarray}
plus
\begin{eqnarray}
h_{A-,A+}^{({\rm hxgn})} = {t_{\perp}^{(0)}\over l} \sum_{n_3=0}^{n-2}\sum_{n_1=0}^{n+1}
(& e^{i[-{\bi k}_i\cdot\delta{\bi r}_{A/A}(-) + {\bi k}_j\cdot\delta{\bi r}_{A/A}(+)]} + \nonumber \\
 & e^{i[-{\bi k}_i\cdot\delta{\bi r}_{A/A}^{\prime}(-) + {\bi k}_j\cdot\delta{\bi r}_{A/A}^{\prime}(+)]} + \nonumber \\
 & e^{i[-{\bi k}_i\cdot\delta{\bi r}_{A/A}^{\prime\prime}(-) + {\bi k}_j\cdot\delta{\bi r}_{A/A}^{\prime\prime}(+)]}) 
\label{h_A-_A+_hxgn}
\end{eqnarray}
if $n > 1$,
with $3$-fold-symmetric displacements from the twist center
\begin{eqnarray}
\delta{\bi r}_{A/A}(\pm) = m_1 {\bi c}_1(\pm) + m_2 {\bi c}_2(\pm) , \nonumber \\
\delta{\bi r}_{A/A}^{\prime}(\pm) = m_1 {\bi c}_2(\pm) + m_2 {\bi c}_3(\pm) , \nonumber \\
\delta{\bi r}_{A/A}^{\prime\prime}(\pm) = m_1 {\bi c}_3(\pm) + m_2 {\bi c}_1(\pm) .
\label{dr_A/A's}
\end{eqnarray}
Again,
the phase factors and minus signs in (\ref{h-+}) have been suppressed above 
in (\ref{h_A-_A+_ncls}) and in (\ref{h_A-_A+_hxgn}).
Also above in (\ref{dr_A/A's}),
 $m_1$ and $m_2$ are integers that satisfy $(m_1 + m_2)$ mod $3 = 1$ on the A sites,
while the integers $n_3$ and $n_1$ denote the column and row, respectively,
 of an A site in the previous parallelogram.
This parallelogram
is the union of the trapezoid adjacent and to the right of the inner triangle shown in Fig. \ref{twist_A_A}
with the outer triangle on top and adjacent to the trapezoid.
The location of an A-site in the parallelogram is (\ref{dr_A/A's})
$\delta{\bi r}_{A/A} = m_1 {\bi c}_1 + m_2 {\bi c}_2$.
The relationship between $(m_1, m_2)$ and the column-row integers $n_3$ and $n_1$ is then
\begin{equation}
(m_1, m_2)\cdot ({\bi c}_1, {\bi c}_2) = (1,-3)\cdot ({\bi c}_1, {\bi c}_2) + (n_3, n_1)\cdot ({\bi a}_3, {\bi a}_1) .
\label{m1_m2_n3_n1}
\end{equation}
Above, $(1,-3)$ are the coordinates for the bottom-left corner of the parallelogram.
Substituting in the identities
${\bi a}_3 = {\bi c}_1 - {\bi c}_2$ and
${\bi a}_1 = {\bi c}_2 - {\bi c}_3 = {\bi c}_1 + 2 {\bi c_2}$ above then yields
\begin{eqnarray}
(m_1, m_2) &=& (1,-3) + n_3 (1,-1) + n_1 (1,2) \nonumber \\
           &=& (1 + n_3 + n_1, -3 - n_3 + 2 n_1).
\label{n3_n1_m1_m2}
\end{eqnarray}
As a consistency check,
notice that the number of terms in the the matrix elements (\ref{h_A-_A+_ncls}) plus (\ref{h_A-_A+_hxgn})
is $3 n (n+1)$, which is equal to $l-1$.

\subsection{$w_0^{\prime}$ and $w_1^{\prime}$}
At in-going and out-going momenta
${\bi k}_j = K_{\widebar{M}} {\hat{\bi x}} = {\bi k}_i$
(see  the red-dot-in-black-circle, Fig. \ref{g_twst_7}),
minus expression (\ref{h_A-_B+}) for the A-B matrix element reduces to
expression (\ref{w_1_prime}) for $w_1^{\prime}$ in the text:
\begin{equation}
w_1^{\prime} = -(t_{\perp}^{(1)}/l) (1 + 6\, {\rm Re}\, S_1^{\prime}),
\label{w'1_S}
\end{equation}
 with series sum
\begin{eqnarray}
S_1^{\prime} &=& \sum_{n_1 = 1}^n \sum_{n_2 = 0}^{n_1 - 1} (e^{i 2\pi /3})^{n_1+n_2} \nonumber \\
    &=& \sum_{n_1 = 1}^n (e^{i 2\pi /3})^{n_1} {1-(e^{i 2\pi /3})^{n_1}\over{1- e^{i 2\pi /3}}} \nonumber \\
    &=& {1\over{1 - e^{i 2\pi /3}}} 
\Biggl[e^{i 2\pi /3} {1-(e^{i 2\pi /3})^{n}\over{1 - e^{i 2\pi /3}}} - {\rm c.c.}\Biggr] .
\label{w'1_gmtrc_sum}
\end{eqnarray}
Above, we have used the result
\begin{eqnarray}
S_{0,m} (z) &=& 1 + z + z^2 + ... + z^m \nonumber \\
            &=& {1-z^{m+1}\over{1-z}}
\label{gmtrc_srs_smmtn}
\end{eqnarray}
for the summation of a geometric series.
After some manipulation, we arrive at the  result
\begin{equation}
S_1^{\prime} = -{1 -\sqrt{3} i\over 3} \Biggl({1\over 2} + \cos\Biggl[{2\pi\over 3}(n-1)\Biggr]\Biggr)
\end{equation}
for the series sum.  Substitution in (\ref{w'1_S}) thereby yields the final result
\begin{equation}
w_1^{\prime} = 2 {t_{\perp}^{(1)}\over l} \cos\Biggl[{2\pi\over 3}(n-1)\Biggr]
\end{equation}
for the matrix element.

Also, at the same in-going and out-going momenta, the A-A matrix elements above,
 (\ref{h_A-_A+_ncls}) and (\ref{h_A-_A+_hxgn}),
reduce to
\begin{equation}
h_{A-,A+}^{({\rm ncls})} = {1\over l} \sum_{(m_1,m_2)=(0,-2),(1,0)} t_{\perp}^{(0)}
\zeta_{A/A}^{(m_1, m_2)} ({\bi k})
\label{h_A-_A+_ncls_bis}
\end{equation}
and
\begin{equation}
h_{A-,A+}^{({\rm hxgn})} = {1\over l} \sum_{n_3=0}^{n-2}\sum_{n_1=0}^{n+1} t_{\perp}^{(0)}
\zeta_{A/A}^{(m_1, m_2)} ({\bi k}) ,
\label{h_A-_A+_hxgn_bis}
\end{equation}
where $\zeta_{A/A}^{(m_1, m_2)} ({\bi k})$ is a $3$-fold-rotation-symmetric phase factor 
given by (\ref{zeta_AA})-(\ref{dc_i}),
with ${\bi k} = K_{\widebar{M}} {\hat{\bi x}}$.
It is shown in the text that the latter reduces to  expression (\ref{zeta_perp_k_m}),
which when substituted into expressions (\ref{h_A-_A+_ncls_bis}) and (\ref{h_A-_A+_hxgn_bis}) above
for $h_{A-,A+}^{({\rm ncls})}$ and $h_{A-,A+}^{({\rm hxgn})}$ 
equal zero because of $3$-fold rotation symmetry.

We must therefore compute the  gradient of 
the phase factor (\ref{zeta_AA}) $\zeta_{A/A}^{(m_1, m_2)} ({\bi k})$
at in-going and out-going momenta ${\bi k}_j = K_{\widebar{M}} {\hat{\bi x}} = {\bi k}_i$:
\begin{eqnarray}
{\bi \nabla}_{{\bi k}} \zeta_{A/A}^{(m_1,m_2)}|_{K_{\widebar{M}} {\hat{\bi x}}} = 
i [& \delta{\bi c}_0(m_1,m_2) e^{i K_{\widebar{M}} {\hat{\bi x}}\cdot\delta{\bi c}_0(m_1,m_2)} + \nonumber \\
   & \delta{\bi c}_+(m_1,m_2) e^{i K_{\widebar{M}} {\hat{\bi x}}\cdot\delta{\bi c}_+(m_1,m_2)} + \nonumber \\
   & \delta{\bi c}_-(m_1,m_2) e^{i K_{\widebar{M}} {\hat{\bi x}}\cdot\delta{\bi c}_-(m_1,m_2)}] .
\label{grdnt_zeta_AA}
\end{eqnarray}
Above, $\delta{\bi c}_0(m_1,m_2)$, $\delta{\bi c}_+(m_1,m_2)$, and $\delta{\bi c}_-(m_1,m_2)$,
and the corresponding dot products with $K_{\widebar{M}} {\hat{\bi x}}$
are given by expressions (\ref{3_fold_c}), (\ref{dc_i}), and (\ref{dot_products_K_M_c}) in the text.
Direct calculation yields the result
\begin{eqnarray}
{\bi \nabla}_{{\bi k}} \zeta_{A/A}^{(m_1,m_2)}|_{K_{\widebar{M}} {\hat{\bi x}}} &=&
3 (\sin \alpha) a  (e^{-i 2\pi/3})^{m_1 - m_2} \cdot \nonumber \\
&& \cdot (e^{+i 2\pi/3} m_1 + e^{-i 2\pi/3} m_2) ({\hat{\bi x}} - i {\hat{\bi y}}).
\label{grdnt_zeta_AA_bis}
\end{eqnarray}
We therefore have the matrix element of the form $w_0 = w_0^{(1)} (\delta k_x - i \delta k_y)$,
with group velocity of the form $w_0^{\prime (1)} = w_{0,{\rm ncls}}^{\prime (1)} + w_{0,{\rm hxgn}}^{\prime (1)}$.
By (\ref{h_A-_A+_ncls_bis}) and (\ref{grdnt_zeta_AA_bis}) above, 
the contribution from the ``nucleus'' to the group velocity is
$w_{0,{\rm ncls}}^{\prime (1)} = 3 (\sin \alpha) a  t_{\perp}^{(0)} / l$.  
Along the Periodic Moir\'e Pattern line, $m = n+1$, we have $2 \sin \alpha = 1/\sqrt{l}$,
and therefore 
$w_{0,{\rm ncls}}^{\prime (1)} = (3/2)  a  t_{\perp}^{(0)} / l^{3/2}$.

Also, by (\ref{h_A-_A+_hxgn_bis}) and (\ref{grdnt_zeta_AA_bis}) above,
the contribution from the ``hexagon'' to the group velocity is
$w_{0,{\rm hxgn}}^{\prime (1)} = - 3 (\sin \alpha) a  t_{\perp}^{(0)} S_0^{\prime} / l$,
with summation
\begin{equation}
S_0^{\prime} = \sum_{n_3=0}^{n-2} \sum_{n_1=0}^{n+1} (e^{i 2 \pi/3})^{m_2 - m_1} (e^{i 2 \pi/3} m_1 + e^{-i 2 \pi/3} m_2) .
\label{S_0_1}
\end{equation}
Next, call $z_1 = e^{-i 2 \pi/3}$ and $z_2 = e^{+i 2 \pi/3}$. Observe that we then have the identity
\begin{eqnarray}
S_0^{\prime} &=& \sum_{n_3=0}^{n-2} \sum_{n_1=0}^{n+1} z_1^{m_1} z_2^{m_2} (z_2 m_1 + z_1 m_2) \nonumber \\
          &=& z_1 z_2 \Biggl({\partial\over{\partial z_1}} + {\partial\over{\partial z_2}}\Biggr)
              \sum_{n_3=0}^{n-2} \sum_{n_1=0}^{n+1} z_1^{m_1} z_2^{m_2} .
\label{S_0_1_bis}
\end{eqnarray}
By (\ref{n3_n1_m1_m2}),
the new summation above acted upon by the partial derivatives is then
\begin{eqnarray}
S_0^{\prime \prime} &=& \sum_{n_3=0}^{n-2} \sum_{n_1=0}^{n+1} z_1^{1 + n_3 + n_1} z_2^{-3 - n_3+ 2 n_1} \nonumber \\
                 &=& z_1^1 z_2^{-3} \sum_{n_3=0}^{n-2} (z_1/z_2)^{n_3} \sum_{n_1=0}^{n+1} (z_1 z_2^2)^{n_1} \nonumber \\
                 &=& z_1 z_2^{-3} \cdot {1-(z_1 / z_2)^{n-1}\over{1-(z_1 / z_2)}} \cdot
                                  {1-(z_1 z_2^2)^{n+2}\over{1-z_1 z_2^2}} .
\label{S_0_prm_1}
\end{eqnarray}
Substituting this expression for $S_0^{\prime \prime}$ into the summation in
expression (\ref{S_0_1_bis}) for $S_0^{\prime}$ yields the result
\begin{eqnarray}
S_0^{\prime} &=& 3 z_1\Biggl({1-z_2^{n-1}\over{1-z_2}}\Biggr)^2 {1-z_1\over{1-z_2}} \nonumber \\
                 && - z_1 {z_2^{n-1} -z_1^{n-1}\over{(1-z_2)^2}} [(n-4) z_2 -(n+5)] .
\label{S_0_prm_1_bis}
\end{eqnarray}
The second term above dominates at large $n$, where we get the limiting result
\begin{equation}
{\rm lim}_{n\rightarrow\infty} S_0^{\prime} = n z_1 (2 i \sin[ 2\pi(n-1)/3] ) / (1-z_2) .
\label{lim_S_0_1}
\end{equation}
Using the identity $1 - z_2 = i \sqrt{3}\, z_1$ thereby yields the result
\begin{equation}
{\rm lim}_{n\rightarrow\infty} S_0^{\prime} = {2\over{\sqrt{3}}} n \sin[ 2\pi(n-1)/3] .
\label{lim_S_0_1_bis}
\end{equation}
In conclusion, at large $n$, the group velocity is given by 
the above ``hexagon''contribution:
%
\begin{equation}
{\rm lim}_{n\rightarrow\infty} w_{0,{\rm hxgn}}^{\prime (1)} = - 3 (\sin \alpha) {a\, t_{\perp}^{(0)}\over l} \cdot 
 {2\over{\sqrt{3}}} n\, \sin \Biggl[{2\pi\over 3} (n-1)\Biggr] .
\label{lim_w_0_1_hxgn}
\end{equation}
Using the limit ${\rm lim}_{n\rightarrow\infty} \sin \alpha = (2 \sqrt{3} n)^{-1}$ for half the twist angle
along the Periodic Moir\'e Pattern line then yields the final result for the group velocity at large $n$:
\begin{equation}
{\rm lim}_{n\rightarrow\infty} w_0^{\prime (1)} = - {a\, t_{\perp}^{(0)}\over l} \sin\Biggl[{2\pi\over 3}(n-1)\Biggr] .
\label{lim_w_0_1}
\end{equation}

\subsection{$w_0$ and $w_1$}\label{ppndx_c_w0_w1}
And at in-going and out-going momenta
${\bi k}_j = M_{M} {\hat{\bi y}} = {\bi k}_i$
(see  Figs. \ref{g_twst_7} and \ref{dgnrt_prtrbtn_M_M}),
minus expression (\ref{h_A-_B+}) for the A-B matrix element reduces to
expression (\ref{w_1}) for $w_1$ in the text:
\begin{equation}
w_1 = -{t_{\perp}^{(1)}\over l}[1
+ 2\, {\rm Re} \sum_{n_1 = 1}^n \sum_{n_2=0}^{n_1 - 1} 
(z_0^{-n_2} + z_0^{n_2-n_1} + z_0^{n_1})],
\label{w_1_bis}
\end{equation}
where $z_0 = {\rm exp} [i (2 \pi /3) \sqrt{3} (\sin \theta)]$.
The sum due to the first term inside the parentheses above is
\begin{eqnarray}
S_1 (1) &=& \sum_{n_1 = 1}^n \sum_{n_2=0}^{n_1 - 1} z_0^{-n_2} \nonumber \\
        &=& n + (n-1) z_0^{-1} (n-2) z_0^{-2} + ... + z_0^{-(n-1)} \nonumber \\
        &=& z_0^{-(n-1)}(1 + 2 z_0 + 3 z_0^2 + ... + n z_0^{n-1}).
\label{S_1_(1)}
\end{eqnarray}
Hence, the sum above is obtained from the derivative of 
the sum $S_{0,n}(z_0)$ of the geometric series (\ref{gmtrc_srs_smmtn})
by
\begin{equation}
S_1 (1) = {1\over{z_0^{n-1}}} {d S_{0,n} \over{d z_0}} ,
\label{S_1_(1)_bis}
\end{equation}
with
\begin{equation}
{d S_{0,n} \over{d z_0}} =  {1\over{(1 - z_0)^2}} [1 - (n+1) z_0^n + n z_0^{n+1}] .
\label{drvtv_gmtric_srs_smmtn}
\end{equation}
Next,
the sum due to the second term inside the parentheses above in (\ref{w_1_bis}) is
\begin{eqnarray}
S_1 (2) &=& \sum_{n_1 = 1}^n  z_0^{-n_1} \sum_{n_2=0}^{n_1 - 1} z_0^{+n_2} \nonumber \\
        &=& \sum_{n_1 = 1}^n  z_0^{-n_1} {1-z_0^{n_1}\over{1-z_0}} \nonumber \\
        &=& {1\over{1-z_0}} \Biggl[{1-z_0^{-(n+1)}\over{1-z_0^{-1}}} - 1 - n\Biggr] .
\label{S_1_(2)}
\end{eqnarray}
Last,
the sum due to the third term inside the parentheses above in (\ref{w_1_bis}) is
\begin{eqnarray}
S_1 (3) &=& \sum_{n_1 = 1}^n  z_0^{n_1} \sum_{n_2=0}^{n_1 - 1} 1 \nonumber \\
        &=& z_0 + 2 z_0^2 + 3 z_0^3 + ... + n z_0^{n} .
\label{S_1_(3)}
\end{eqnarray}
It is related to the derivative (\ref{drvtv_gmtric_srs_smmtn}) of
the sum $S_{0,n}(z_0)$ of the geometric series (\ref{gmtrc_srs_smmtn}) by 
\begin{equation}
S_1 (3) = z_0 {d S_{0,n}\over{d z_0}} .
\label{S_1_(3)_bis}
\end{equation}
%
It is $z_0^n$ times the series sum (\ref{S_1_(1)_bis}) $S_1 (1)$ from the first term.

The final sum in $w_1$ above (\ref{w_1_bis}) is $2\, {\rm Re}[S_1(1) + S_1(2) + S_1(3)]/l$.
It can be evaluated in simple form at large $n$.
In particular,
from (\ref{sine}) and (\ref{l_M}), we identify  the useful limits
\begin{equation}
l\rightarrow 3 \Biggl(n+{1\over 2}\Biggr)^2 \quad {\rm and} \quad z_0 \rightarrow {\rm exp} \Biggl({i\over{n+{1\over 2}}} {2\pi\over 3}\Biggr)
\label{lrg_n}
\end{equation}
as $n$ approaches infinity.
Plugging in these limits into the expressions above then yields the final result for the matrix element:
\begin{equation}
{\rm lim}_{n\rightarrow\infty} w_1 = - {3\over{2\pi}}\Biggl({1\over{\sqrt{3}}} + {3\over{2\pi}}\Biggr) t_{\perp}^{(1)}
\cong - 0.5036\, t_{\perp}^{(1)} .
\label{w_1_lim_bis}
\end{equation}
Note that
it is important here to plug in 
the large-$n$ limit $1-z_0 \rightarrow - (i/n)(2\pi /3)$ in the denominators of the above expressions,
(\ref{S_1_(1)_bis}), (\ref{drvtv_gmtric_srs_smmtn}), (\ref{S_1_(2)}) and (\ref{S_1_(3)_bis}), 
for the sums $S_1(1)$, $S_1(2)$, and $S_1(3)$.

Last, at the same in-going and out-going momenta, the A-A matrix elements above,
 (\ref{h_A-_A+_ncls}) and (\ref{h_A-_A+_hxgn}),
reduce to expressions (\ref{h_A-_A+_ncls_bis}) and (\ref{h_A-_A+_hxgn_bis})
 in terms of the $3$-fold rotation symmetric
phase factor (\ref{zeta_AA})-(\ref{dc_i}) $\zeta_{A/A}^{(m_1, m_2)}({\bi k})$,
but evaluated at momentum ${\bi k} = M_M {\hat{\bi y}}$ instead.
It is given by expression (\ref{zeta_perp_m_m}) in the text.
Then by (\ref{h_A-_A+_ncls_bis}),
we get the following contribution to the A-A matrix element from the ``nucleus'':
\begin{equation}
w_{0, {\rm ncls}} = - t_{\perp}^{(0)} (3 z_0^{2/3} + 2 z_0^{-1/3} + z_0^{-4/3}) / l .
\label{w_0_ncls}
\end{equation}
Also, by (\ref{h_A-_A+_hxgn_bis}) above and (\ref{zeta_perp_m_m}) in the text,
the contribution from the ``hexagon'' to the A-A matrix element has the form
$w_{0, {\rm hxgn}} = - t_{\perp}^{(0)} S_0 / l$, with sum
\begin{equation}
S_0 = \sum_{n_3=0}^{n-2} \sum_{n_1=0}^{n+1} 
[(z_0^{1/3})^{2 m_1 - m_2} + (z_0^{1/3})^{-(m_1 + m_2)} + (z_0^{1/3})^{2 m_2 - m_1}] .
\label{S_0}
\end{equation}
By (\ref{n3_n1_m1_m2}), the exponents above are:
\begin{eqnarray}
2 m_1 - m_2 &=& 5 + 3 n_3 , \nonumber \\
m_1 + m_2   &=& -2 + 3 n_1 , \nonumber \\
2 m_2 - m_1 &=& -7 - 3 n_3 + 3 n_1 .
\label{xpnnts_S_0}
\end{eqnarray}
The sum due to the first term in (\ref{S_0}) is then
\begin{eqnarray}
S_0 (1) &=& \sum_{n_3=0}^{n-2} \sum_{n_1=0}^{n+1} z_0^{5\over 3} z_0^{n_3} \nonumber \\
        &=& (n+2) z_0^{5/3} {1-z_0^{n-1}\over{1-z_0}},
\label{S_0(1)}
\end{eqnarray}
the sum due to the second term in (\ref{S_0}) is then
\begin{eqnarray}
S_0 (2) &=& \sum_{n_3=0}^{n-2} \sum_{n_1=0}^{n+1} z_0^{{2\over 3}} z_0^{-n_1} \nonumber \\
        &=& (n-1) z_0^{2/3} {1-z_0^{-(n+2)}\over{1-z_0^{-1}}},  
\label{S_0(2)}
\end{eqnarray}
and the sum due to the third term in (\ref{S_0}) is then
\begin{eqnarray}
S_0 (3) &=& z_0^{-7/3} \sum_{n_3=0}^{n-2} z_0^{-n_3} \sum_{n_1=0}^{n+1} z_0^{n_1} \nonumber \\
        &=& z_0^{-7/3} {1-z_0^{-(n-1)}\over{1-z_0^{-1}}} {1-z_0^{n+2}\over{1-z_0}} .
\label{S_0(3)}
\end{eqnarray}
Figure \ref{angle_vs_n} in the text demonstrates that the present matrix element
$w_{0, {\rm ncls}} +w_{0, {\rm hxgn}}$
takes the form $w_0 = w_0^{(0)} e^{i \alpha}$ at large $n$,
where $w_0^{(0)}$ is real.  Also, direct calculation of
$S_0 = S_0(1) + S_0(2) + S_0(3)$
yields that the large-$n$ limit
for the present matrix element, ${\rm lim}_{n\rightarrow\infty} w_0$,
 coincides with the corresponding limit (\ref{w_1_lim_bis}) for $w_1$,
but with the replacement $t_{\perp}^{(1)} \rightarrow t_{\perp}^{(0)}$.
Last, notice that the previous phase factor $e^{i \alpha}$ clearly 
contains the first-order correction by the twist angle to the matrix element $w_0$ 
in the limit of large $n$.

\end{document}